\documentclass[sn-mathphys-ay]{sn-jnl}

\usepackage{graphicx}%
\usepackage{multirow}%
\usepackage{amsmath,amssymb,amsfonts}%
\usepackage{amsthm}%
\usepackage{mathrsfs}%
\usepackage[title]{appendix}%
\usepackage{xcolor}%
\usepackage{textcomp}%
\usepackage{manyfoot}%
\usepackage{booktabs}%
\usepackage{algorithm}%
\usepackage{algorithmicx}%
\usepackage{algpseudocode}%
\usepackage{listings}%

\newcommand{\bx}{\boldsymbol x}

\newcommand{\bX}{\boldsymbol X}

\newcommand{\indep}{\rotatebox[origin=c]{90}{$\models$}}

\usepackage{array}
\newcolumntype{C}[1]{>{\centering\arraybackslash}m{#1}}

\usepackage{tikz-cd}    
\usetikzlibrary{backgrounds}
\usetikzlibrary{positioning}
\usepackage{tikz}
\usetikzlibrary{arrows,automata} 

\usetikzlibrary{calc}
\usepackage[font={small}]{caption}

\theoremstyle{definition}

\newtheorem*{notation*}{Notation and conventions}

\begin{document}

\title[Loglinear modelling of huge contingency tables]{Loglinear modelling of huge contingency tables}


\author*[1]{\fnm{Veronica} \sur{Vinciotti}}\email{veronica.vinciotti@unitn.it}

\author[2]{\fnm{Ernst C.} \sur{Wit}}\email{ernst.jan.camiel.wit@usi.ch}

\affil*[1]{\orgdiv{Department of Mathematics}, \orgname{University of Trento}, \orgaddress{\street{Via Sommarive 14}, \city{Trento}, \postcode{38123}, \state{(TN)}, \country{Italy}}}

\affil[2]{\orgdiv{Institute of Computing}, \orgname{Universit\`a della Svizzera italiana}, \orgaddress{\street{Via la Santa 1}, \city{Lugano}, \postcode{6962}, \state{Ticino}, \country{Switzerland}}}

\abstract{Contingency tables are the canonical representation of multivariate categorical data. As the size of the contingency table grows exponentially with the number of variables, even a moderate number of variables, each with a moderate number of levels, results in a huge number of cells, the majority of which remains empty even with a significant amount of data. We propose efficient methods for inferring higher-order loglinear models  by performing subsampling on the set of the empty cells. 
First, we derive the likelihood under a zero-deflated Poisson sampling scheme. This is maximized via an efficient iteratively re-weighted least squares algorithm, leading to consistent and close to efficient estimators. This method works well for moderately sized contingency tables, but runs into computational instability when the number of dimensions grows. By sacrificing some efficiency, we show that nested case-control multinomial sampling combined with a degenerate logistic regression approach is also consistent and can be applied to arbitrarily large contingency tables.
We illustrate the method with an analysis of data from the General Social Survey, which consists of $15014$ observations in a $69$-dimensional contingency table with a total of $6.6\times 10^{38}$ cells.}


\maketitle

\section{Introduction}\label{sec:intro}
Multivariate categorical data are collected in many application fields, such as social, behavioral, financial, and biomedical sciences.  This type of data is represented in the form of a contingency table, where each cell reports the
counts associated to a combination of levels of the variables. Loglinear models provide a general class of models to analyze these data in order to find statistical dependences between the variables \citep{agresti12}. \textcolor{black}{In the context of hierarchical log-linear models for contingency tables under Poisson or multinomial sampling schemes, the log-likelihood function is concave, and the maximum likelihood estimator exists and is unique if and only if the observed marginal totals corresponding to the minimal sufficient statistics are strictly positive. The likelihood equations equate these observed marginals to their expected counterparts, involving only the marginal tables defined by the generating class of the hierarchical model \citep{birch1963maximum, bishop1975discrete} and an iteratively proportional fitting algorithm. However, for sparse contingency tables the MLE is unlikely to exist, making this classical approach unsuitable \citep{nardi12}.}

Statistical inference for loglinear models is challenging already for a moderate number of categorical variables, as there are both  statistical and computational issues that are in part unique to this type of data. Firstly, the discreteness of the data means that a large number of parameters is needed to describe complex relationships between the variables. Secondly, a moderate number of variables and/or the presence of variables with many categories leads to a large number of combinations of levels and therefore to prohibitively large contingency tables. With finite sample sizes, many of these combinations will be filled with zero counts. 

A number of approaches have addressed these problems using methods that, in different ways, reduce the huge contingency table into lower dimensional tables which are then used for approximate inferential procedures. In particular, \cite{dahinden10} approximate the joint distribution of the variables with a factorization over a suitably defined decomposable graph and perform loglinear modelling within each lower dimensional clique of this representation. \cite{aliverti22} restrict the attention to models with main effects and two-way interaction terms and propose an approximation of the joint likelihood with a composite likelihood over bivariate contingency tables.  Despite strong sparsity inducing priors, this Bayesian inferential procedure  has a high computational burden. As an alternative, \cite{dobra18} approximate the joint likelihood  via a pseudo-likelihood of conditional distributions of a node given its neighbours, which are then modelled non-parametrically. They propose a stochastic search algorithm through the space of loglinear graphical models within a Bayesian inferential procedure and provide an implementation of this for the case of binary categorical variables in the R package \texttt{BDgraph} \citep{bdgraph}. The approach is computationally efficient for structural learning and does not impose restrictions on the highest order of the interactions, but does not return estimates of the loglinear effects.

As an alternative to the approaches above, penalized likelihood has been proposed for parameter estimation of loglinear models. Since the dependence between any two variables is represented by a set of parameters, which are all zero in the case of conditional independence \citep{roverato17},  group lasso allows to perform model selection for complex loglinear models \citep{dahinden10,nardi12}. In this case, the highest order of the interactions must be imposed a priori and the selected model is not guaranteed to be graphical. Moreover, while addressing the high dimensionality of the parameter space, these penalized approaches cannot handle the computational challenge of extremely large sample sizes. For the specific case of  contingency tables with ordinal categorical variables, alternative penalized approaches can be developed using discrete Gaussian copula graphical models  \citep{behrouzi2019detecting, mohammadi17}.

In this paper, we offer an alternative solution to these problems and propose a method that is able to estimate parameters of loglinear models of generic orders  from huge contingency tables. Based on the fact that most of the information is contained in the cells with positive counts,  
we propose to use  only a random sample of cells with zero counts. 
\textcolor{black}{Under a Poisson sampling scheme for the contingency table, this leads to a zero-deflated Poisson likelihood. In section~\ref{sec:mediumlarge}, we derive the likelihood conditional on the random sampling of zeros and develop an efficient iteratively re-weighted least squares algorithm for its maximization. A simulation study shows how the estimators are  consistent and close to efficient. This method works well for moderately sized contingency tables, but runs into computational instability when the number of dimensions grows and the contingency table becomes very sparse. By sacrificing some efficiency, we show, in section~\ref{sec:huge}, that nested case-control multinomial sampling combined with a degenerate logistic regression approach is also consistent and can be applied efficiently to arbitrarily large contingency tables using standard implementations of logistic regression models. Both the zero-deflated Poisson and the degenerate logistic approaches can be easily augmented with a ridge regularization. This is particularly useful in the case of high-dimensional sparse contingency tables where the maximum likelihood estimator of the effects is unlikely exist \citep{fienberg12,nardi12}.}

We conclude the paper with  an illustration on multivariate categorical survey data. In particular, in section~\ref{sec:realdata}, we study structural dependences among different cultural dimensions by considering 69 questions from the General Social Survey, having 3-5 levels each \citep{bertrand23}. The questions  generate a contingency table of size $6.6 \times 10^{38}$, of which $15014$ are non-empty cells with one count each. This setting, which is common for modern applications of loglinear models, aligns well with the challenges discussed above and provides the ideal scenario for the illustration of the proposed methods. 

\section{Inference of medium-large contingency tables}
\label{sec:mediumlarge}
We start this section by recalling the full likelihood of a loglinear model under a Poisson sampling scheme. Let $\bX=(X_1,\ldots,X_p)$ be a multivariate categorical random vector taking values in $\mathcal{L}_1 \times \cdots \times \mathcal{L}_p$. This leads to a contingency table of size $|\mathcal{L}_1|\times \ldots \times |\mathcal{L}_p|$. Let $\bx$ be a cell of the contingency table. The rate of $\bX$ taking this particular value can be parametrized via a loglinear expansion \citep{roverato17},
\begin{equation}
\log \mu_{\bx} = \sum_{D \subseteq V} \lambda_D(\bx_D),
\label{eq:loglin}
\end{equation}
with $V=\{1,\ldots,p\}$ and functions $\lambda_D:\prod_{j\in D}\mathcal{L}_j \longrightarrow \mathbb{R}$ that depend only on $\bx_D$. These functions  are uniquely identifiable, if one adds, for example, a set-first-to-zero constraint, i.e., $\lambda_D(\bx_D)=0$ whenever $x_j=l_{j1}$ for any $j \in D$. The terms with $|D|=1$ are referred to as main effects, those with $|D|=2$ as two-way interactions, and so forth. 

A specific loglinear model includes only a subset of terms of this expansion, while the other terms are set to zero. Setting specific sets of parameters to zero results in conditional independence structures among the variables \citep{lauritzen96}. In particular, assuming $p(\bx)>0$, and considering only models where functions $\lambda_{D'}$ with $D \subseteq D'$ are zero when $\lambda_D$ is zero, i.e., models that are hierarchical, then the association between the loglinear expansion and conditional independence between variables simplifies to
\begin{equation}
X_i \: \indep \: X_j | \bX_{V\backslash\{i,j\}} \Longleftrightarrow \lambda_{\{i,j\}}(x_i,x_j)= 0  \quad \forall (x_i,x_j) \in \mathcal{L}_i \times \mathcal{L}_j,
\label{eq:condind}
\end{equation}
as a direct consequence of the Hammersley-Clifford theorem.

\subsection{Poisson sampling scheme}
Statistical inference of loglinear models is about estimation of the parameters of the corresponding loglinear expansion as well as recovery of the underlying conditional independence graph. Let  $\boldsymbol{n}=\{n_{\bx}\}$ be a $p$-way contingency table of dimension $|\mathcal{L}_1|\times \ldots \times |\mathcal{L}_p|$ generated by a model \eqref{eq:loglin}. Under a Poisson sampling scheme,  the counts $N_{\bx}$ are independent and identically distributed with $E[N_{\bx}]=\mu_{\bx}$, i.e.,
\[ P(N_{\bx}=n_{\bx})= \frac{\mu_{\bx}^{n_{\bx}}}{n_{\bx}!}e^{-\mu_{\bx}}.\]
Since $\mu_{\bx}=e^{\sum_{D \subseteq V}\lambda_D(\bx_D)}$,
we can rewrite the rate as
\[ \mu_{\bx}=e^{\boldsymbol{\lambda}^{\top}\boldsymbol{m}_{\bx}},\]
where the vector $\boldsymbol{m}_{\bx}$ has length equal to the total number of non-vanishing $\lambda$ terms in the model, and has entries equal to 1 corresponding to the $\lambda$ terms that depend on the levels $\bx$ only. 
This generative model leads to the log-likelihood
\begin{equation}
\ell^{\text{poi}}_{\boldsymbol{n}}(\boldsymbol{\lambda})=\log\prod_{\bx \in \mathcal{L}_1 \times \cdots \times \mathcal{L}_p} P(N_{\bx}=n_{\bx})\propto\sum_{\bx \in \mathcal{L}_1 \times \cdots \times \mathcal{L}_p} \Big(-\mu_{\bx}+n_{\bx}\log(\mu_{\bx})\big).
\label{eq:full-likelhood}
\end{equation}

Let $M$ be the design matrix with rows $\boldsymbol{m}_{\bx}$. This matrix has as many columns as the number of parameters in the loglinear expansion of the model under consideration and as many rows as the number of cells of the contingency table. Both dimensions can be very large, as they are intrinsically connected. Considering the case of $p=69$ categorical variables with $k=4$ levels each, similar to the illustration in section~\ref{sec:realdata}, a model with all main effects and two-way interaction terms but no higher-order terms leads to a design matrix with $k^p=4^{69} \approx 3.5 \times 10^{41}$ rows and $1+(k-1)p+(k-1)^2\dbinom{p}{2}=21322$ columns. The expectation is that the underling generative model is associated to a sparse conditional independence graph, i.e., that a number of these parameters is zero.  In particular, using \eqref{eq:condind},  a missing edge between two variables would correspond to a set of $(k-1)^2=9$ parameters being zero.

\subsection{Subsampling empty cells}
While group lasso approaches can guarantee both the existence of the estimators and the sparsity of the underlying conditional independence graph \citep{nardi12}, they do not resolve the issue of the super-exponential number of rows of the contingency table.  In this paper, we solve this issue  by considering a sample  $\mathcal{S}$ of cells from the full contingency table. If the contingency table is very large, such as in the example described above, random sampling leads most surely to a sample of empty cells. A different strategy is needed for the construction of $\mathcal{S}$.

Clearly, one would like to select the most informative cells for the estimation of the parameters $\boldsymbol{\lambda}$.  In particular, from the theory of generalized linear models \citep{McCullagh89}, the asymptotic variance-covariance matrix of $\boldsymbol{\hat\lambda}$ is given by $({M}^{\top}UM)^{-1}$ with  $U$ the diagonal matrix with diagonal elements given by the variances $\mu_{\bx}$. The subset of sampled cells should then be selected so as to minimize the variance-covariance matrix of $\boldsymbol{\hat\lambda}$ or, alternatively, to maximize its inverse, the information matrix. Since the diagonal of the matrix $U$ gives a higher weight to rows of $M$ with high expected counts $\mu_{\bx}$, the cells with positive counts, representing the actual observations, should be part of $\mathcal{S}$ as they carry most of the information. For the remaining cells, we propose a random sampling with probability $\pi$. In particular, we consider
\[
 N_{\bx} \sim \mbox{Poi}\left(e^{\boldsymbol{\lambda}^{\top}\boldsymbol{m}_{\bx}}\right), \quad
P(\bx \in \mathcal{S} \:|\: N_{\bx}) =\begin{cases} 1 & N_{\bx} \ge 1 \\ \pi & N_{\bx}=0\end{cases}.
\]

As in \cite{wang22}, we now consider the likelihood of the data conditional on our chosen sampling strategy. In particular, since the probability of any cell being sampled is given by
\begin{align*}
P(\bx \in \mathcal{S}) &= P(\bx \in \mathcal{S}| N_{\bx} \ge 1)P(N_{\bx} \ge 1)\!+\!P(\bx \in \mathcal{S}| N_{\bx} =0)P(N_{\bx} =0) \\
&= 1-e^{-\mu_{\bx}}+\pi e^{-\mu_{\bx}} = 1-(1-\pi)e^{-\mu_{\bx}},
\end{align*}
the \textcolor{black}{zero-deflated Poisson loglikelihood} is given by
\begin{align*}
\ell^{\text{zdpoi}}_{\mathcal{S}}(\boldsymbol{\lambda})&=\sum_{\substack{\bx \in \mathcal{S} \\ n_{\bx} \ge 1}} \log\Big(e^{-\mu_{\bx}}\frac{{\mu_{\bx}}^{n_{\bx}}}{n_{\bx}!}\frac{1}{1-(1-\pi)e^{-\mu_{\bx}}}\Big) + \sum_{\substack{\bx \in \mathcal{S} \\ n_{\bx}=0}} \log\Big(\frac{\pi e^{-\mu_{\bx}}}{1-(1-\pi)e^{-\mu_{\bx}}}\Big). 
\end{align*}
Since $n_{\bx}=0$ for all elements in the second summation, and ignoring additive constants, this can be rewritten as
\begin{align}
\ell^{\text{zdpoi}}_{\mathcal{S}}(\boldsymbol{\lambda})&
= \sum_{\bx \in \mathcal{S}}\Big(-\mu_{\bx}+n_{\bx}\log(\mu_{\bx})-\log(1-(1-\pi)e^{-\mu_{\bx}})\Big).
\label{eq:sample-likelhood}
\end{align}
As expected, for $\pi=1$ the sampled data corresponds to the full contingency table and the likelihood is simply that of a Poisson. On the other hand, for values of $\pi$ different to $1$,  the generative process accounts for the fact that positive counts are retained while empty cells are randomly sampled, down to the extreme of $\pi=0$ where the sampled data contain only the positive counts and the likelihood is that of a Poisson truncated at $0$.

\subsection{Zero-deflated Poisson estimators of loglinear effects}
In this section, we derive the zero-deflated Poisson likelihood estimators of the loglinear effects and discuss their properties.
Consider a contingency table $\boldsymbol{n}=\{n_{\bx}\}$, a loglinear model with $\log(\mu_{\bx})= \boldsymbol{\lambda}^{\top}\boldsymbol{m}_{\bx}$,  and a sample $\mathcal{S}$ of the contingency table generated with the strategy described above for a fixed $\pi \in [0,1]$. The estimator of the parameters  $\boldsymbol{\lambda}$ is found by maximizing $\ell^{\text{zdpoi}}_{\mathcal{S}}(\boldsymbol{\lambda})$ in \eqref{eq:sample-likelhood}. 
Since the first two terms of \eqref{eq:sample-likelhood} correspond to a Poisson likelihood, while
\[
\frac{\partial}{\partial\boldsymbol{\lambda}} \log(1-(1-\pi)e^{-\mu_{\bx}}) = \frac{(1-\pi)\mu_{\bx}e^{-\mu_{\bx}}}{1-(1-\pi)e^{-\mu_{\bx}}}\boldsymbol{m}_{\bx},
\]
the estimator $\hat{\boldsymbol{\lambda}}$ solves the likelihood score equations
\begin{align}
\frac{\partial \ell^{\text{zdpoi}}_{\mathcal{S}}}{\partial\boldsymbol{\lambda}}&= \sum_{\bx \in \mathcal{S}} (n_{\bx}-\mu_{\bx})\boldsymbol{m}_{\bx}-\sum_{\bx \in \mathcal{S}}\frac{(1-\pi)\mu_{\bx}e^{-\mu_{\bx}}}{1-(1-\pi)e^{-\mu_{\bx}}}\boldsymbol{m}_{\bx}=0,
\label{eq:score-likelihood}
\end{align}
with $\mu_{\bx}=e^{\boldsymbol{\lambda}^{\top}\boldsymbol{m}_{\bx}}$.

Taking the derivative of the score function in equation~\eqref{eq:score-likelihood}, the expected Fisher information matrix is given by
\begin{align}
-\frac{\partial}{\partial\boldsymbol{\lambda}}\frac{\partial \ell^{\text{zdpoi}}_{\mathcal{S}}}{\partial\boldsymbol{\lambda}}&= -\frac{\partial}{\partial\boldsymbol{\lambda}}\Big(\sum_{\bx \in \mathcal{S}} (n_{\bx}-\mu_{\bx})\boldsymbol{m}_{\bx}-\sum_{\bx \in \mathcal{S}}\frac{(1-\pi)\mu_{\bx}e^{-\mu_{\bx}}}{1-(1-\pi)e^{-\mu_{\bx}}}\boldsymbol{m}_{\bx}\Big)=M_{\mathcal{S}}^{\top}WM_{\mathcal{S}},
\label{eq:fisherinformation}
\end{align}
where $M_{\mathcal{S}}=(\boldsymbol{m}^{\top}_{\bx})_{\bx \in \mathcal{S}}$ is the design matrix associated to the sampled cells  $\mathcal{S}$ and $W$ is a diagonal matrix with diagonal elements  
\[W_{\bx}=\mu_{\bx}+(1-\pi)\mu_{\bx}\frac{e^{\mu_{\bx}}-\mu_{\bx}e^{\mu_{\bx}}-(1-\pi)}{(e^{\mu_{\bx}}-(1-\pi))^2}.\]
The inverse of this matrix, evaluated at $\hat{\boldsymbol{\lambda}}$, gives the asymptotic variance-covariance matrix of the estimators, i.e.,
\[V(\hat{\boldsymbol{\lambda}})=(M_{\mathcal{S}}^{\top}WM_{\mathcal{S}})^{-1}.\]
Notice how the diagonal of the matrix $W$ gives a higher weight to rows of $M_{\mathcal{S}}$ with high expected counts $\mu_{\bx}$. As discussed before, the cells with positive counts, i.e., those with a large $\mu_{\bx}$,  are associated to a higher information, and thus to a lower variance.

\subsection{IRWLS algorithm for parameter estimation} \label{sec:irwls}
In this section, we discuss computational aspects related to obtaining zero-deflated Poisson likelihood estimates of loglinear effects. In particular, we derive an efficient algorithm for solving the likelihood score equations~\eqref{eq:score-likelihood}. By setting \[\mu^*_{\bx}=\mu_{\bx} \Big( 1+\frac{(1-\pi)e^{-\mu_{\bx}}}{1-(1-\pi)e^{-\mu_{\bx}}}\Big),\]
these can be conveniently rewritten as 
\begin{align*}
\frac{\partial \ell^{\text{zdpoi}}_{\mathcal{S}}}{\partial\boldsymbol{\lambda}} = \sum_{\bx \in \mathcal{S}} (n_{\bx}-\mu^*_{\bx})\boldsymbol{m}_{\bx}=0.
\end{align*}
or, in matrix notation,
\begin{align}
M_{\mathcal{S}}^{\top}(\boldsymbol{n}_{\mathcal{S}}-\boldsymbol{\mu}^*)=0,
\label{eq:score-equations}
\end{align}
with $\boldsymbol{n}_{\mathcal{S}}$ the vector of counts associated to the sampled cells $\mathcal{S}$ and $\boldsymbol{\mu}^*$ the associated vector of $\mu^*_{\bx}$.
Since
\[\log\mu^*_{\bx}=\log\mu_{\bx} +\log\Big( 1+\frac{(1-\pi)e^{-\mu_{\bx}}}{1-(1-\pi)e^{-\mu_{\bx}}}\Big),\] a first approach is to view the score equations as those of a Poisson generalized linear model with an offset given by $\log\Big( 1+\dfrac{(1-\pi)e^{-\mu_{\bx}}}{1-(1-\pi)e^{-\mu_{\bx}}}\Big)$. 
However, since this term depends on the parameters, an iterative procedure based on this may be numerically unstable. A more stable solution is to rewrite the problem as an iteratively reweighted least-squares approach, similar to that used for generalized linear models \citep{McCullagh89}.

To this end, using the score function in~\eqref{eq:score-equations} and the expected Fisher information matrix from~\eqref{eq:fisherinformation}, the Fisher scoring update at iteration $t$ is given by
\[
\boldsymbol{\lambda}^{(t+1)}= \boldsymbol{\lambda}^{(t)}+(M_{\mathcal{S}}^{\top}WM_{\mathcal{S}})^{-1}M_{\mathcal{S}}^{\top}(\boldsymbol{n}_{\mathcal{S}}-{\boldsymbol{\mu}^*}),
\]
with $W$ and ${\boldsymbol{\mu}^*}$ evaluated at $\boldsymbol{\lambda}^{(t)}$.
This can be equivalently rewritten as
\[
\boldsymbol{\lambda}^{(t+1)}= \boldsymbol{\lambda}^{(t)}+(M_{\mathcal{S}}^{\top}WM_{\mathcal{S}})^{-1}M_{\mathcal{S}}^{\top}W W^{-1}(\boldsymbol{n}_{\mathcal{S}}-{\boldsymbol{\mu}^*}).
\]
Multiplying both sides by $M_{\mathcal{S}}^{\top}WM_{\mathcal{S}}$, we get
\begin{align*}
M_{\mathcal{S}}^{\top}WM_{\mathcal{S}}\boldsymbol{\lambda}^{(t+1)} &= M_{\mathcal{S}}^{\top}WM_{\mathcal{S}}\boldsymbol{\lambda}^{(t)}+M_{\mathcal{S}}^{\top}W W^{-1}(\boldsymbol{n}_{\mathcal{S}}-{\boldsymbol{\mu}^*}) \\
&= M_{\mathcal{S}}^{\top}W(M_{\mathcal{S}}\boldsymbol{\lambda}^{(t)}+W^{-1}(\boldsymbol{n}_{\mathcal{S}}-{\boldsymbol{\mu}^*})
\\
&= M_{\mathcal{S}}^{\top}W(\boldsymbol{\eta}^{(t)}+W^{-1}(\boldsymbol{n}_{\mathcal{S}}-{\boldsymbol{\mu}^*}).
\end{align*}
These are the estimated equations of a weighted least squares with a working response defined by
\begin{equation}
\boldsymbol{z}=\boldsymbol{\eta}^{(t)}+W^{-1}(\boldsymbol{n}_{\mathcal{S}}-{\boldsymbol{\mu}^*}), 
\label{eq:workingresponse}
\end{equation}
and adjusting the current linear predictor $\boldsymbol{\eta}^{(t)}$, and weights defined by the matrix $W$. The weighted least-squares solution provides then the next update of the parameters 
\[
\boldsymbol{\lambda}^{(t+1)}=(M_{\mathcal{S}}^{\top}WM_{\mathcal{S}})^{-1}M_{\mathcal{S}}^{\top}W\boldsymbol{z},
\]
and the algorithm is iterated until convergence. \textcolor{black}{In the empirical analyzes, we declare convergence when the largest absolute score is lower than $0.005$. As for initial values, we set $\mu_{\bx}=n_{\bx}+0.5$, as in the traditional IRWLS algorithm for generalized linear models. } 

\textcolor{black}{Although computationally efficient unique estimates of the maximum likelihood estimator of $\boldsymbol{\lambda}$ may exist based on collapsed counts in the contingency table, in general this is not the case for sparse high-dimensional contingency tables.
In these cases, a standard ridge penalty can be added to the objective 
function, i.e.,}
\begin{align*}
\ell^{\text{zdpoi}}_{\mathcal{S},\rho}(\boldsymbol{\lambda})&
= \sum_{\bx \in \mathcal{S}}\Big(-\mu_{\bx}+n_{\bx}\log(\mu_{\bx})-\log(1-(1-\pi)e^{-\mu_{\bx}})\Big) -\rho {\boldsymbol{\lambda}}^{\top}{\boldsymbol\lambda}.
\end{align*}
This leads to the penalized score
\begin{align*}
\frac{\partial \ell^{\text{zdpoi}}_{\mathcal{S},\rho}}{\partial \boldsymbol{\lambda}}=M_{\mathcal{S}}^{\top}(\boldsymbol{n}_{\mathcal{S}}-\boldsymbol{\mu}^*)-2\rho\boldsymbol{\lambda},
\label{eq:penalized-score-equations}
\end{align*}
the expected Fisher information matrix
\[
-\frac{\partial}{\partial\boldsymbol{\lambda}}\frac{\partial \ell^{\text{zdpoi}}_{\mathcal{S},\rho}}{\partial\boldsymbol{\lambda}}=M_{\mathcal{S}}^{\top}WM_{\mathcal{S}}+2\rho I,
\]
and thus a Fisher scoring update
\[
\boldsymbol{\lambda}^{(t+1)}= \boldsymbol{\lambda}^{(t)}+(M_{\mathcal{S}}^{\top}WM_{\mathcal{S}}+2\rho I)^{-1}(M_{\mathcal{S}}^{\top}(\boldsymbol{n}_{\mathcal{S}}-\boldsymbol{\mu}^*)-2\rho\boldsymbol{\lambda}^{(t)}).
\]
Following similar steps to before,  this leads to the update
\[
\boldsymbol{\lambda}^{(t+1)}=(M_{\mathcal{S}}^{\top}WM_{\mathcal{S}}+2\rho I)^{-1}M_{\mathcal{S}}^{\top}W\boldsymbol{z},
\]
with the working response $\boldsymbol{z}$ defined as before by equation~\eqref{eq:workingresponse}.

\subsection{\textcolor{black}{Comparison of zero-deflated and full Poisson estimates}} \label{sec:simulation1}

In this section, we perform a simulation study by generating data from high-dimensional and sparse contingency tables. The aim of the study is to assess the performance of the zero-deflated Poisson likelihood estimates when we consider only a small fraction of randomly sampled zero cells.

We consider the case of $p=13$ categorical variables with $3$ levels each. Thus, the contingency table has size $3^{13}$. We further consider a two-way interaction model, therefore a total of $339$ parameters. We draw the values of the $\lambda$ effects from a $\rm{Beta}(0.25,0.25)$ distribution. We further subtract $0.5$, so that the values are centered at zero, and set the intercept to $\log(339)-13\log(3)$, i.e., roughly one count per parameter for a model with only an intercept. When we consider the $\lambda$ values simulated as above, and calculate $\sum_{\bx}\mu_{\bx}=\sum_{\bx} \exp(\boldsymbol{\lambda}^{\top}\boldsymbol{m}_{\bx})$, we obtain an expected count of $4764.85$. This means that only about $0.3\%$ of the cells of the contingency table are expected to contain some counts.

\begin{figure}[t]
     \centering
    \begin{minipage}[b]{0.45\linewidth}
        \centering        \includegraphics[width=\linewidth]{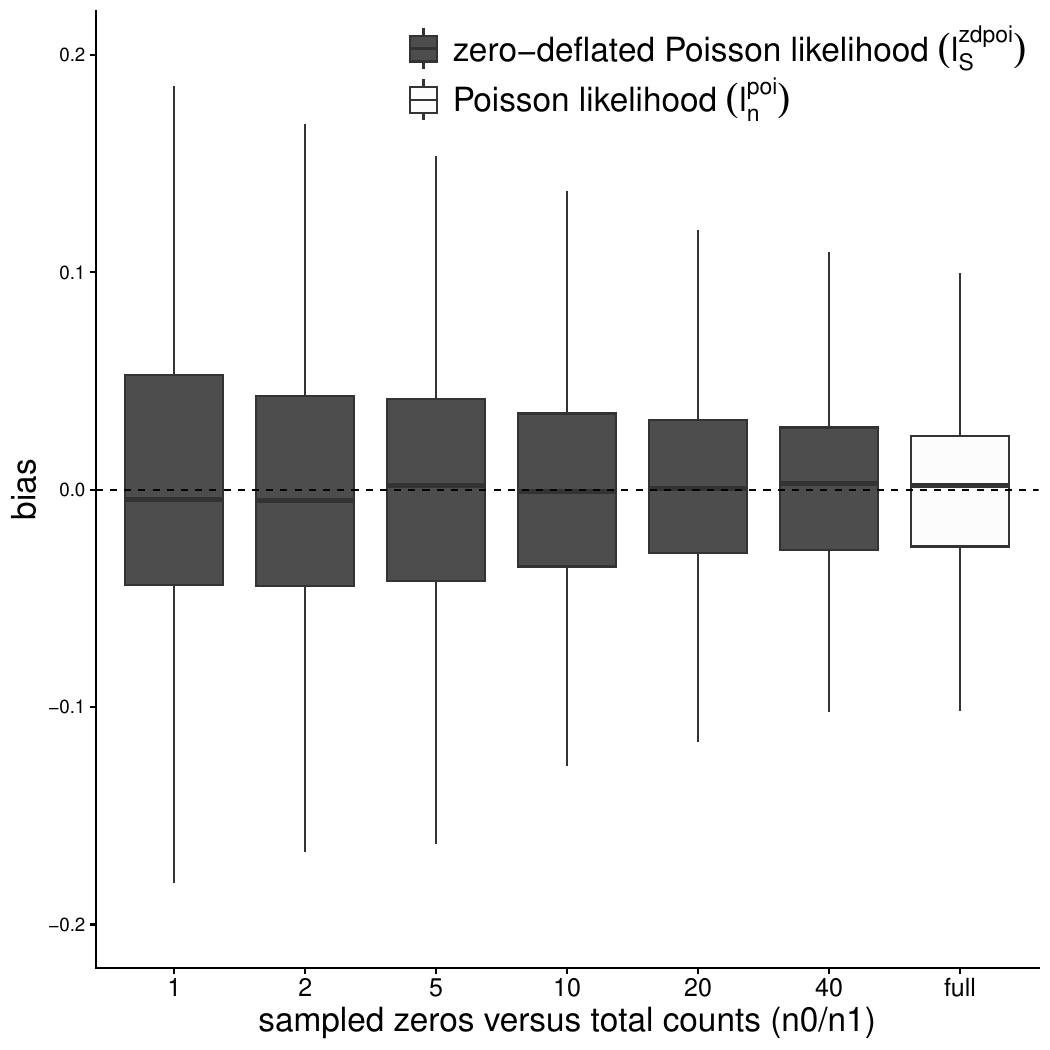}\\
(a)    
\end{minipage}
    \hfill
    \begin{minipage}[b]{0.45\linewidth}
        \centering
        \includegraphics[width=\linewidth]{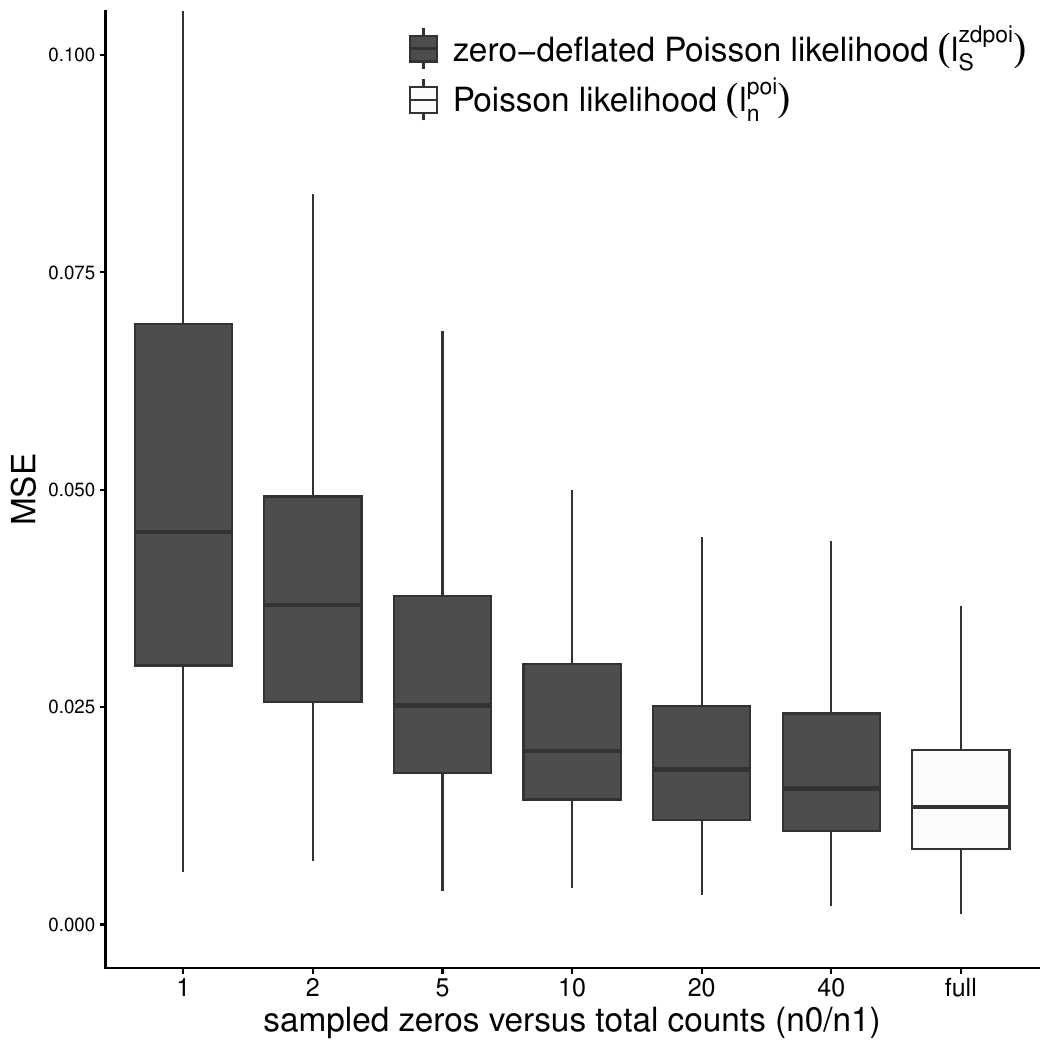}\\
(b)
    \end{minipage}  \hfill
      \caption{Simulation with $p=13$ categorical variables with $3$ levels each from a two-way model. Comparison  between zero-deflated Poisson  under random sampling of zeros and   Poisson likelihood estimates from the full contingency table, in terms of (a) bias  and (b) MSE of the estimated interaction effects $\boldsymbol{\lambda}$.  The estimators are approximately unbiased and close to efficient already for a small number of empty cells ($n_0$) compared to the total counts ($n_1$). }
    \label{fig:sim_pi}
\end{figure}

We generate $10$ datasets from the model described above.  We then consider $6$ scenarios, where we take the observed counts in each dataset and  add a number of randomly sampled empty cells ($n_0$) equal to $1, 2, 5, 10, 20, 40$ times the total counts in each dataset ($n_1$), respectively. In this way, we create samples of the full contingency table that contain a small fraction $\pi$ of empty cells, ranging from about $0.3\%$ ($n_0=n_1$) up to $12\%$ ($n_0=40n_1$). \textcolor{black}{For each of the $6$ settings, Figure~\ref{fig:sim_pi} shows boxplots of the bias and mean squared error (MSE)  across the $338$ estimates of the $\boldsymbol{\lambda}$ parameters, excluding the intercept which is on a different scale.}
\textcolor{black}{The bias and MSE of each parameter are calculated across the $10$ replicated datasets. 
We compare estimation via the IRWLS approach of section~\ref{sec:irwls}, based on the  zero-deflated Poisson likelihood $\ell^{\text{zdpoi}}_{\mathcal{S}}$ from equation~\eqref{eq:sample-likelhood},  with the estimates from the Poisson likelihood $\ell^{\text{poi}}_{\textbf{n}}$ from equation~\eqref{eq:full-likelhood} using the full contingency table.
The results show how the estimates from the zero-deflated Poisson likelihood are approximately unbiased and closed to efficient already for a small fraction of empty cells compared to the total.} 

\section{\textcolor{black}{Inference of huge contingency tables}} \label{sec:huge}

	\textcolor{black}{The techniques described in the previous section work well for small to medium sized contingency
	tables. However, a moderate number of categorical variables can lead to huge contingency tables. In the case 
	of  $p = 69$ categorical variables with $k = 4$ levels each, the contingency table is of size $4^{69} \approx 3.5 \times 10^{41}$.  Even a
	two-way loglinear model requires a design matrix with $4^{69}$ rows, making direct
	maximization of the full Poisson likelihood impossible.  Sampling of zeros, as in the previous section, would reduce the size of the design matrix. However, the huge size of the contingency table means that moderately-sized datasets can only be attained by sampling a tiny percentage $\pi$ of empty cells. Moreover, the probability of counts in any cell, that is $1-e^{-\mu_{\bx}}$, is also close to zero, creating issues in the calculation of zero-deflated Poisson likelihood~\eqref{eq:sample-likelhood} with any achievable numerical precision.
	The key insight discussed in Section~\ref{sec:mediumlarge} is that, in a sparse
	contingency table, the vast majority of cells are empty and the non-empty cells
	carry most of the statistical information.  In this section we develop a parallel
	treatment of the huge-table regime under a \emph{multinomial} sampling scheme, which
	conditions on the total observed counts $n_1 = \sum_{\bx} n_{\bx}$.  We first derive the
	full conditional likelihood (Section~\ref{sec:multinomial}), show why it is
	intractable for huge tables, and then propose two complementary subsampling
	strategies in Section~\ref{sec:conditional_logistic}.}
	
\textcolor{black}{	\subsection{Multinomial sampling scheme}}
	\label{sec:multinomial}

    \textcolor{black}{Under the multinomial sampling scheme, the total counts $n_1 = \sum_{\bx} n_{\bx}$ is treated as fixed 
    \citep{agresti12}.  Conditional on the total, the cell counts
	$\{N_{\bx}\}$ follow a multinomial distribution with probabilities
	\begin{equation}
		p_{\bx}(\boldsymbol{\lambda}) = \frac{\mu_{\bx}}{\sum_{\bx'} \mu_{\bx'}}
		= \frac{e^{\boldsymbol{\lambda}^{\top} \boldsymbol{m}_{\bx}}}{\sum_{\bx'} e^{\boldsymbol{\lambda}^{\top} \boldsymbol{m}_{\bx'}}},
		\label{eq:mult_probs}
	\end{equation}
	where $\mu_{\bx} = e^{\boldsymbol{\lambda}^{\top} \boldsymbol{m}_{\bx}}$ is the Poisson rate from the loglinear
	expansion~\eqref{eq:loglin} and the normalization sum runs over all cells
	$\bx' \in \mathcal{L}_1 \times \cdots \times \mathcal{L}_p$.  The multinomial
	log-likelihood is
	\begin{eqnarray}
		\ell^{\text{mult}}_{\boldsymbol{n}}(\boldsymbol{\lambda})
		&=& \sum_{\bx\,:\,n_{\bx} \geq 1} n_{\bx} \log p_{\bx}(\boldsymbol{\lambda}) \nonumber\\
		&=& \sum_{\bx\,:\,n_{\bx} \geq 1} n_{\bx} \boldsymbol{\lambda}^{\top} \boldsymbol{m}_{\bx}
		- n_1 \log \Bigl(\sum_{\bx' \in \mathcal{L}_1 \times \cdots \times \mathcal{L}_p} e^{\boldsymbol{\lambda}^{\top} \boldsymbol{m}_{\bx'}}\Bigr) + C.
    \label{eq:mult_lik}
	\end{eqnarray}
	Since $p_{\bx}$ depends on $\boldsymbol{\lambda}$ only through the loglinear expansion, the
	multinomial and Poisson likelihoods share the same score equations and, consequently,
	the same maximum likelihood estimator $\hat{\boldsymbol{\lambda}}$, except for the intercept, which is
	unidentified under the multinomial scheme.  The asymptotic covariance
	matrix of $\hat{\boldsymbol{\lambda}}$ under the multinomial scheme is
	\begin{equation}
		V^{\text{mult}}(\hat{\boldsymbol{\lambda}})
		= \bigl(M^{\top} \tilde{U} M\bigr)^{-1},
		\label{eq:mult_var}
	\end{equation}
	where $\tilde{U}$ is the diagonal matrix with entries $\tilde{u}_{\bx} = n_1 p_{\bx} (1-p_{\bx})$
	for a two-category table and, more generally, its multinomial analogue \citep{McCullagh89}.}
	
\textcolor{black}{	The computational bottleneck of~\eqref{eq:mult_lik} is the normalization constant,
	which requires a sum over all $|\mathcal{L}|=\prod_{j=1}^{p}|\mathcal{L}_j|$ cells.
    If the maximum likelihood estimator exists, then algorithms based on the summary statistics 
    of sub-tables of the full contingency table can be used within an iterative fitting procedure. 
    However, as mentioned before, already for a model of moderate complexity, within a high dimensional setting, the MLE is unlikely
    to exist \citep{fienberg12,nardi12}. Instead, we propose subsampling strategies in combination with regularization
    to obtain robust estimates for the entire sample space. }
	
\textcolor{black}{		\subsection{Subsampling of empty cells under a multinomial scheme}
	\label{sec:conditional_logistic}
    In this section, we introduce a subsampling scheme of empty cells, with which we associate three different likelihoods.}
	
\textcolor{black}{	\paragraph{Conditional logistic likelihood}}	
\textcolor{black}{	We retain all $n_1$ observed counts and sample a random set $\mathcal{S}_0$ of $n_0$
	empty cells, giving the sampled set $\mathcal{S} = \mathcal{S}_1 \cup \mathcal{S}_0$. In order to simplify the notation, we now consider $\mathcal{S}_1$ as the set of cases rather than the cells, so $|\mathcal{S}| = n_1 + n_0$. Under the multinomial scheme, the exact conditional
	likelihood conditions on the event that exactly $n_1$ cases occurred among the
	$n_1 + n_0$ total samples. This gives the conditional logistic likelihood of
	\begin{equation}
		L_{\mathcal{S}}(\boldsymbol{\lambda})
		= \frac{\exp\Bigl(\sum_{\bx \in \mathcal{S}_1} 
        \,\boldsymbol{\lambda}^{\top} \boldsymbol{m}_{\bx}\Bigr)}
		{\displaystyle\sum_{A\,\in\,\mathcal{C}(n_1+n_0,\,n_1)}
			\exp\Bigl(\sum_{\bx \in A} \boldsymbol{\lambda}^{\top} \boldsymbol{m}_{\bx}\Bigr)},
		\label{eq:clogit_exact}
	\end{equation}
	where $\mathcal{C}(n_1+n_0, n_1)$ denotes the collection of all subsets of size
	$n_1$ from the $n_1 + n_0$ sampled cells \citep{Prentice1978}.}
	
\textcolor{black}{	The denominator of~\eqref{eq:clogit_exact} sums over $\binom{n_1+n_0}{n_1}$ terms,
	which is computationally intractable for any realistic $n_1$. A standard
	approximation, due to \citet{Breslow1974}, replaces the exact subset sum with the
	$n_1$-th power of the sum of individual contributions, yielding the approximate
	conditional log-likelihood
	\begin{equation}
		\ell^{\text{clogit}}_{\mathcal{S}}(\boldsymbol{\lambda})
		= \sum_{\bx \in \mathcal{S}_1} 
        \,\boldsymbol{\lambda}^{\top} \boldsymbol{m}_{\bx}
		- n_1 \log\Bigl(\sum_{\bx' \in \mathcal{S}} e^{\boldsymbol{\lambda}^{\top} \boldsymbol{m}_{\bx'}}\Bigr),
		\label{eq:clogit_approx}
	\end{equation}
	which is essentially the multinomial likelihood~\eqref{eq:mult_lik} on the sample $\mathcal{S}$. This is the default in standard software implementations such as \texttt{clogit} in the
	\texttt{survival} package \citep{Therneau2000}. Maximizing~\eqref{eq:clogit_approx} 
    is consistent and asymptotically normal as $n_1
	\to \infty$. The standard implementation has a per-iteration cost 
    of  
    $O\!\bigl((n_1 + n_0) q^2\bigr)$, where $q$ is the number of parameters in the model,
	owing to the full Hessian required by the standard Cox-model algorithm. This is similar to the
    zero-deflated Poisson Poisson method introduced in section~\ref{sec:mediumlarge}. In that case, $n_1$ is replaced by the number of non-empty cells, which is  close to  $n_1$ for high-dimensional contingency tables. 
    The algorithm can be slow even for moderate $n_1$ whenever $n_0$ is large relative to $n_1$.  
    The nested case-control sampling scheme
	introduced next resolves this by splitting the single stratum 
    approach of the conditional logistic into $n_1$ small, matched, strata.}

\textcolor{black}{\paragraph{Nested case-control likelihood}
We now introduce a sampling scheme that is computationally more attractive and that
	mirrors the \emph{nested case-control} design of epidemiology
	\citep{Thomas1977,Oakes1981}.  For each case $\bx \in \mathcal{S}_1$ we
	independently draw a random set $\mathcal{C}(\bx)$ of $m$ empty cells, 
    sampling uniformly without replacement from
	$\mathcal{L} \setminus \mathcal{S}_1$. 
	Conditioning on the matched structure, the probability for a single case $\bx$
	with control set $\mathcal{C}(\bx)$ is given by \citep{Goldstein1992}
	\begin{equation*}
		p_{\bx}(\boldsymbol{\lambda})
		= \frac{e^{\boldsymbol{\lambda}^{\top} \boldsymbol{m}_{\bx}}}{e^{\boldsymbol{\lambda}^{\top} \boldsymbol{m}_{\bx}}
			+ \sum_{\bx' \in \mathcal{C}(\bx)} e^{\boldsymbol{\lambda}^{\top} \boldsymbol{m}_{\bx'}}},
		\label{eq:ncc_stratum}
	\end{equation*}
	and the nested case-control log-likelihood is
	\begin{equation}
		\ell_{\mathcal{S}}^{\text{ncc}}(\boldsymbol{\lambda})
		= \sum_{\bx \in \mathcal{S}_1} 
        \left[
		\boldsymbol{\lambda}^{\top} \boldsymbol{m}_{\bx}
		- \log\Bigl(e^{\boldsymbol{\lambda}^{\top} \boldsymbol{m}_{\bx}}
		+ \sum_{\bx' \in \mathcal{C}(\bx)} e^{\boldsymbol{\lambda}^{\top} \boldsymbol{m}_{\bx'}}\Bigr)
		\right].
		\label{eq:ncc_lik}
	\end{equation}
	For arbitrary $m$, the log-likelihood~\eqref{eq:ncc_lik} decomposes into
	$|\mathcal{S}_1|=n_1$ independent strata, each of size $m + 1$.  The per-iteration cost
	of Newton-Raphson is therefore 
    $O\!\bigl(n_1 m q^2\bigr)$: linear in $n_1$,
	linear in $m$, and quadratic in the number of parameters $q$.  For $m = 1$ this
	reduces to $O(n_1 q^2)$.  
    Compared to the $O\!\bigl((n_1 + n_0) q^2\bigr)$
	cost of the direct conditional logistic approach, the nested case-control scheme offers a
	substantial computational saving whenever $n_0$ is large relative to $n_1$.  This
	matches the efficiency of the nested case-control design in survival
	analysis \citep{Langholz1996}.}

\textcolor{black}{\paragraph{Degenerate logistic likelihood}
If the number of empty cell controls is reduced to $m=1$, the conditional probability of $\bx$ can be written as
	\begin{equation*}
		p_{\bx}(\boldsymbol{\lambda})
		= \frac{e^{\boldsymbol{\lambda}^{\top} (\boldsymbol{m}_{\bx}-\boldsymbol{m}_{\bx'})}}{1+e^{\boldsymbol{\lambda}^{\top} (\boldsymbol{m}_{\bx}-\boldsymbol{m}_{\bx'})}
}.
		\label{eq:logistic}
	\end{equation*}
When the number of controls $m$ is greater than one, the likelihood that repeats each case $m$ times will result in consistent estimates of the model parameters $\boldsymbol{\lambda}$. We define this likelihood as follows, 
	\begin{equation}
    \ell_{\mathcal{S}}^{\text{dlogit}}(\boldsymbol{\lambda})
		= \sum_{\bx \in \mathcal{S}_1} 
        \sum_{i=1}^m \left[
		\boldsymbol{\lambda}^{\top} (\boldsymbol{m}_{\bx}-\boldsymbol{m}^{(i)}_{\bx'})
		- \log\Bigl(1+e^{\boldsymbol{\lambda}^{\top} (\boldsymbol{m}_{\bx}-
        \boldsymbol{m}^{(i)}_{\bx'})}\Bigr)
		\right],
		\label{eq:dlogit_lik}
	\end{equation}
with $\boldsymbol{m}^{(i)}_{\bx'}$, $i=1,\ldots,m$, indicating the $m$ controls for the case $\bx$.
This expression happens to coincide with the likelihood of a degenerate logistic likelihood with a fixed response of $1$ and a design matrix corresponding to the difference between the rows of the model matrices of the cases and controls. We denote this matrix by $\Delta M_{\mathcal{S}}$. Since this matrix has $n_1m$ rows and $q$ columns, the asymptotic computational cost is still in the order of $O\!\bigl(n_1 m q^2\bigr)$. However, as we will see later in the simulation study,  this approach is significantly faster in practice. For small $m$, the design matrix $\Delta M_{\mathcal{S}}$ has many fewer rows compared to the one in the nested case-control likelihood $M_{\mathcal{S}}$, whereas for large $m$, the logistic likelihood is a simpler function to optimize.}

\textcolor{black}{It is important to note that for this likelihood, except for $m=1$, the asymptotic variance does not correspond to the inverse of its second derivative. However, as the approach can be seen as an application of a composite likelihood, convenient corrections exist using a sandwich estimator \citep{varin2011overview},
\[V(\hat {\boldsymbol{\lambda}}^{{\tt{ dlogit}}}) = H^{-1} J H^{-1}.\]
Here $H$ is the usual Hessian matrix for the logistic regression, i.e.,
$H= \Delta M^t_\mathcal{S}W\Delta M_{\mathcal{S}}$, where  the diagonal elements of  the $n_0\times n_0$ diagonal matrix $W$ correspond to the probabilities ${\bf p}\circ({\bf 1}_{n_0}-{\bf p})$, and where $J$ is the squared gradient matrix $J = GG^t$, with $G=\Delta M_{\mathcal{S}}^t({\bf 1}_{n_0}-{\bf p})$.}

\textcolor{black}{\paragraph{Ridge penalization and existence of the estimator}
In principle, the same problem of the non-existence of MLEs in a high-dimensional setting applies also to any of the likelihoods introduced in this section. As discussed in section~\ref{sec:irwls}, also the regression framework introduced in this section allows for efficient regularization approaches. In particular, we can easily apply a standard ridge penalty to any of the likelihoods,
\[  \ell_{\mathcal{S},\rho}^{\square}(\boldsymbol{\lambda}) = \ell_{\mathcal{S},\rho}^{\square}(\boldsymbol{\lambda}) - \rho \boldsymbol{\lambda}^\top\boldsymbol{\lambda}, \]
where, in practice, $\square$ stands for either the {\tt ncc} or {\tt dlogit} likelihoods. The tuning parameter is selected via generalized cross-validation.} 

\textcolor{black}{\subsection{Comparison of subsampled and full multinomial estimates}
In this section, we perform a simulation study to evaluate the performance of the subsampling strategies derived in the previous section under a multinomial sampling scheme. Similar to the simulation study under a Poisson sampling scheme (section~\ref{sec:simulation1}), we consider $p=13$ variables with $3$ levels each and a sparse scenario where we fix the total counts to $n_1=3986$, i.e., 0.025\% of the total number of cells of the contingency table. We consider again a two-way interaction model with $\lambda$ effects drawn from a Beta distribution specified as before. The large, though not huge, size of the contingency table means that we can still generate samples from the full contingency table under the different sampling strategies, and compare the estimators obtained via the different methods with the true value of the parameters. Perhaps surprisingly, for larger contingency tables our proposed inference scheme is still computationally feasible (as it scales with the number of counts in the contingency table, not its dimension), but it becomes computationally challenging to \emph{sample} data from such contingency tables for a given model. }

\begin{figure}[t]
     \centering
    \begin{minipage}[b]{0.45\linewidth}
        \centering        \includegraphics[width=\linewidth]{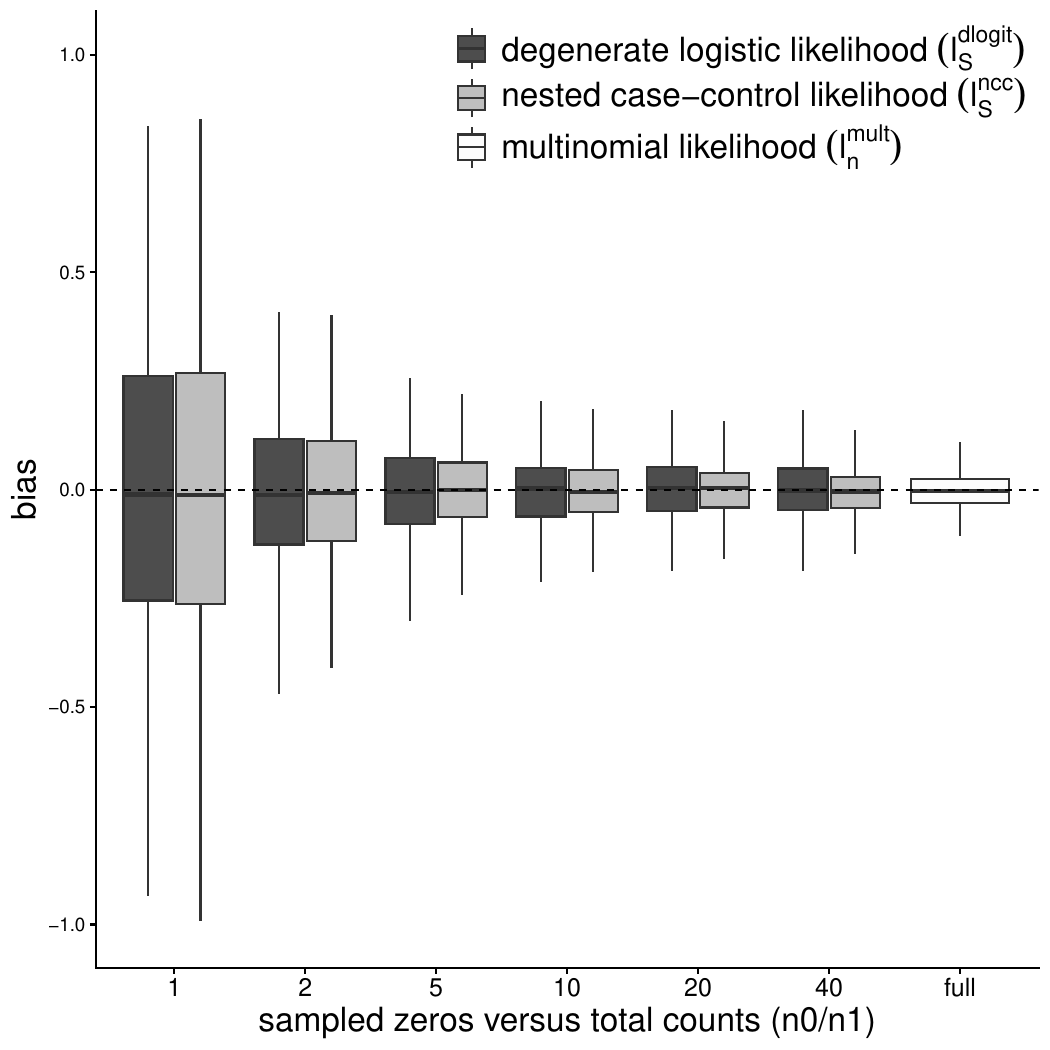}\\
(a)    
\end{minipage}
    \hfill
    \begin{minipage}[b]{0.45\linewidth}
        \centering
        \includegraphics[width=\linewidth]{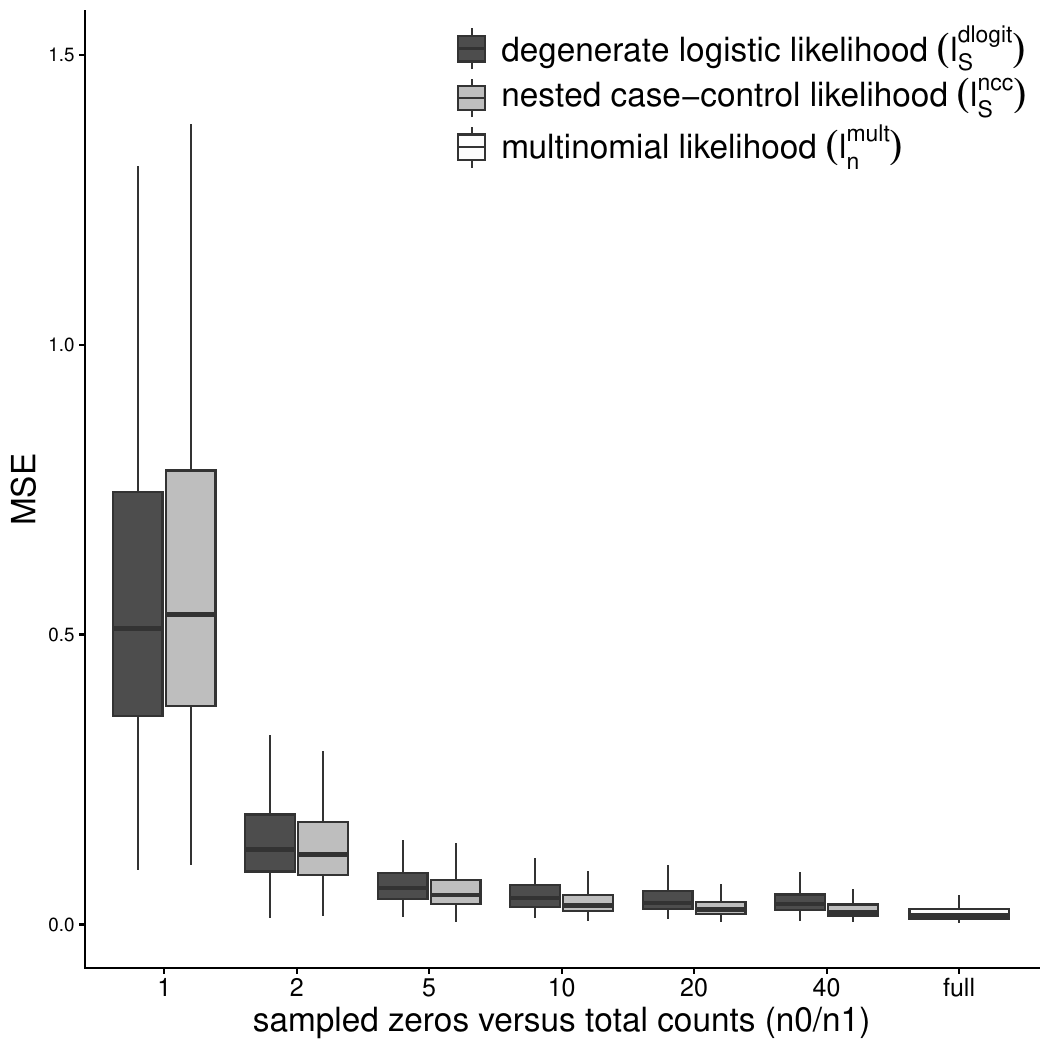}\\
(b)
    \end{minipage}  \hfill
      \caption{Simulation with $p=13$ categorical variables with $3$ levels each from a two-way model. Comparison  between  conditional logistic  from nested case-control sampled data,  its approximation using logistic regression, and   multinomial likelihood estimates from the full contingency table, in terms of (a) bias  and (b) MSE of the estimated interaction effects $\boldsymbol{\lambda}$.  The estimators are approximately unbiased, but have a large variance for a small number of empty cells ($n_0$) compared to the total counts ($n_1$).  }
    \label{fig:sim_clogit}
\end{figure}

\textcolor{black}{Figure~\ref{fig:sim_clogit} reports the results in terms of bias and MSE for the two proposed methods under strata of varying size. In particular, the first method considers the conditional logistic estimates, obtained by maximizing the log-likelihood $\ell_{\mathcal{S}}^{\text{ncc}}$ given in equation~\eqref{eq:ncc_lik} via the function \texttt{clogit}  of the \texttt{survival} R package with option {\tt approximate} \citep{Therneau2000}. The second method considers the approximation via the degenerate logistic regression, i.e., maximizing the log-likelihood $\ell_{\mathcal{S}}^{\text{dlogit}}$ given in equation~\eqref{eq:dlogit_lik}, across the same settings. As a benchmark, the figure includes also bias and MSE of the estimators from the full multinomial likelihood \eqref{eq:mult_lik}. Figure~\ref{fig:sim_clogit}a shows how the estimates from both methods are approximately unbiased, while Figure~\ref{fig:sim_clogit}b shows how the biggest contribution to the MSE is given by the variance of the estimator. This is high for the case of one control per case for both methods (median MSE of 0.53 and 0.51 for nested case-control and degenerate logistic, respectively), while it drops drastically when two controls are used per case (0.12 and 0.13, respectively). As the number of controls increases, the nested case-control estimator approached the variance of the full likelihood estimator already for a moderate number of controls, while the use of degenerate logistic induces a small loss in efficiency compared to the full likelihood also for a large number of controls (median MSE of degenerate logistic equal to 0.04 versus 0.02 of full likelihood with $40$ controls per case). The latter is due to the correlation among the observations where the same case is matched with different controls, which is not accounted for by the logistic regression model.}  

\textcolor{black}{The large variance of the estimators for a small number of controls is instead due to numerical instability in the presence of a large number of parameters. Although the design matrices of both controls and cases  have at least one count in each column across all replicates and settings, the sparse setting of the simulation means that for some datasets there may be a low number of counts in some sub-tables leading to a poor estimation of some effects. This aspect is likely to deteriorate as the dimension of the table increases, but can be resolved with the use of regularization. For the degenerate logistic method,  Figure~\ref{fig:ridge}a shows how the variance is significantly reduced if one uses a ridge penalty, which is here optimized with the use of cross-validation via the \texttt{glmnet} implementation in R \citep{glmnet}.
\begin{figure}[t]
    \centering
       \begin{minipage}[b]{0.45\linewidth}
        \centering
        \includegraphics[width=\linewidth]{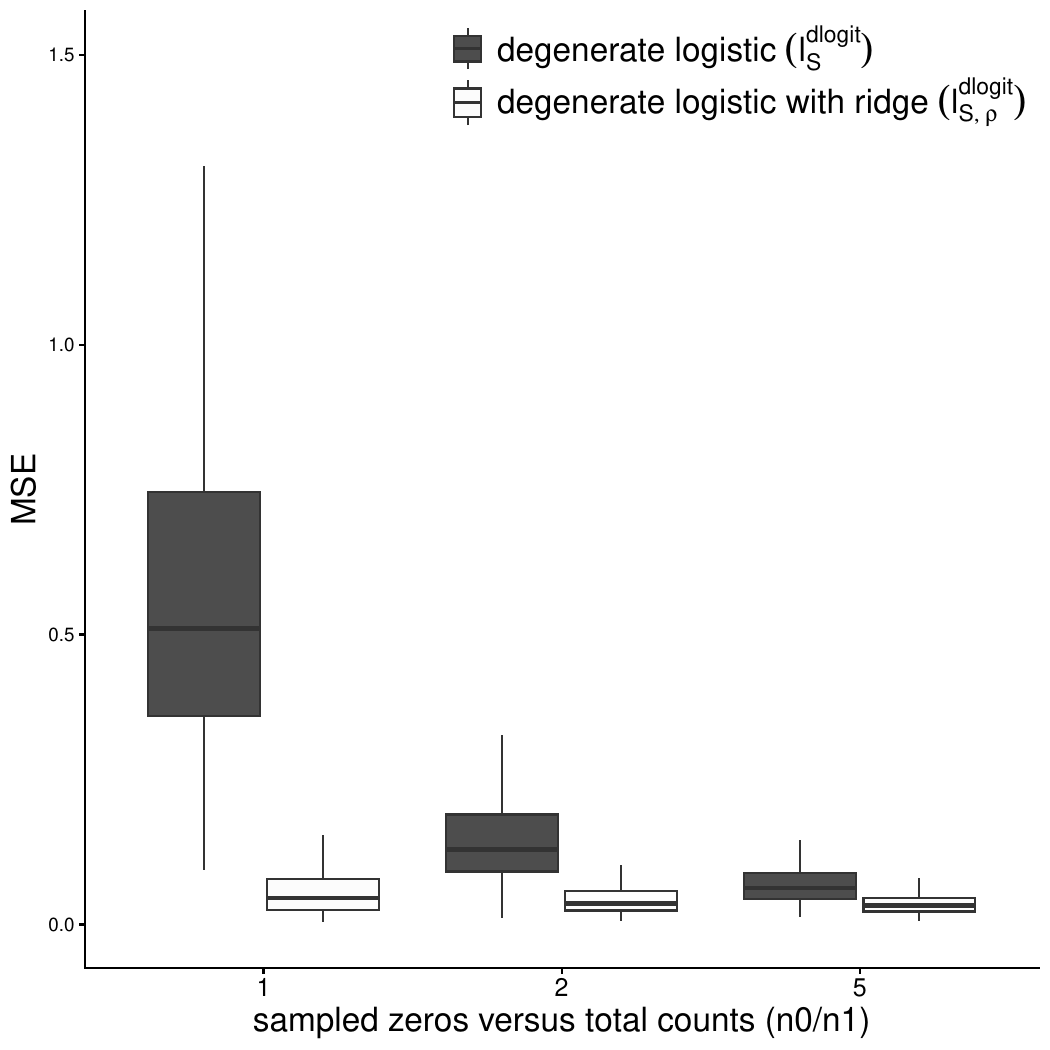}\\
(a)
    \end{minipage}  \hfill
      \begin{minipage}[b]{0.45\linewidth}
        \centering
        \includegraphics[width=\linewidth]{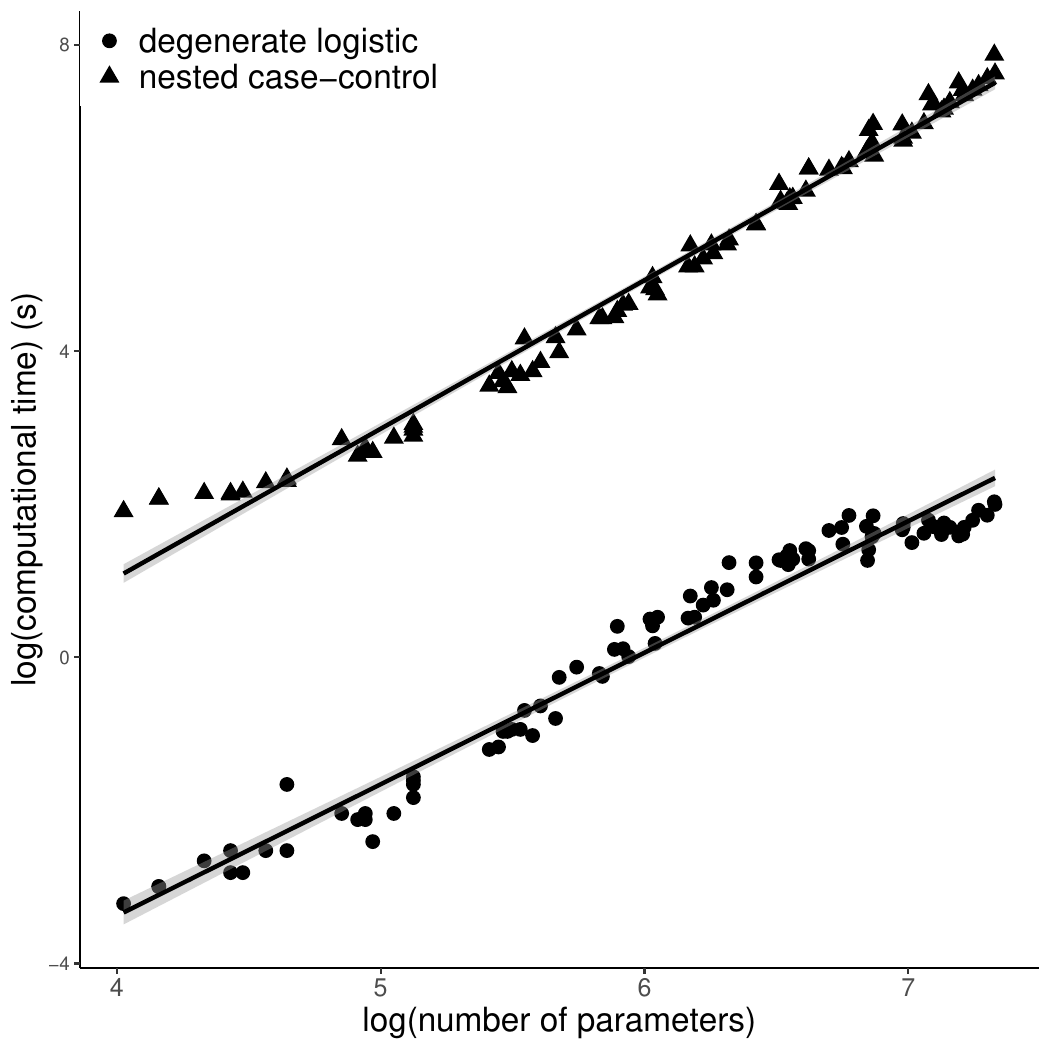}\\
(b)
    \end{minipage}  \hfill
    \caption{Multinomial subsampling methods on huge contingency tables: (a) MSE of estimated effects of a two-way model from data on p=13 categorical variables with 3 levels each is significantly lower if the degenerate logistic effects are estimated with a ridge penalty (b) As $p$ varies between $20$ and $100$, setting $n_1=4000$ and $n_0/n_1=2$, the computational cost of fitting one loglinear graphical model increases nearly quadratically with the number of parameters in the model (slope of log-log plot equal to 1.793 for degenerate logistic regression and 1.941 for nested case-control approach), but degenerate logistic regression is significantly faster.}
    \label{fig:ridge}
\end{figure}}

\textcolor{black}{Finally, we consider the case of huge contingency tables and compare the computational cost of the degenerate logistic approach with that based on the nested case-control likelihood. To this end, we vary $p \in \{20,30,40,50,60,70,80,90, 100\}$ and consider again $3$ levels per categorical variable. Already with $p=20$, the contingency table, of size $3^{20}$ is beyond the limit of current implementations for the fitting of loglinear models, e.g., the popular implementations of \texttt{dmod}   and \texttt{loglm} in R \citep{hoisgaard12} can only handle contingency tables of size up to $2^{31}$. Given the huge size of the contingency table, we cannot sample cases or controls from the full contingency table model, as  in the previous simulations. Instead, we draw multivariate data from a Gaussian copula graphical model with $2$ cutoff points for each marginal. The conditional independence graph of the latent multivariate Gaussian is set to a random structure with 5\% of edges. This generates multivariate ordinal data which we treat as multivariate categorical data. We generate $n_1=4000$ counts as described above and then add $n_0=2n_1$ empty cells by drawing randomly a level for each categorical variable. We fit loglinear graphical models corresponding to the latent Gaussian graph across $10$ replicates of each setting. We  set a small ridge penalty of $\rho=2$ for computational stability. Indeed, as $p$ increases, we notice various instances where either the design matrix of controls or that of cases has a number of columns that are all zero, with the largest case of $72$ columns for one of the simulations when $p=100$. }

\textcolor{black}{Figure~\ref{fig:ridge}b shows the computational advantages of the degenerate logistic compared to the nested case-control approach. Although the computational cost of fitting one loglinear model via degenerate logistic or nested case-control scale both approximately quadratically with the number of parameters in the model (the slopes of the log-log plots are equal to 1.793 and 1.941, respectively), the vertical separation by 5 means that the former is approximately 150 times faster than the latter. }

\section{Model selection} We now turn to the problem of  conducting model selection from a contingency
table dataset.
\textcolor{black}{Likelihood-based  criteria can be used for performing model selection among the space of zero-deflated Poisson or nested case-control likelihoods, depending on the sampling regime considered. In this way, also the model selection step is consistent with the modelling choice made and accounts for the sampling of zeros.} A number of options are available, similar to loglinear model selection approaches from the full contingency table \citep{hoisgaard12}. In particular, we propose a stepwise procedure through graph space: at each step an existing edge is removed or a new edge is added based on some information criterion that penalizes model complexity,
\[IC({\cal M})=-2\ell^{\square}_{\mathcal{S}}(\hat{\boldsymbol{\lambda}})+\kappa\rm{df}({\cal M}),\]
with ${\cal M}$ a candidate loglinear graphical model having $\rm{df}({\cal M})$ number of  parameters $\boldsymbol{\lambda}$ and fitted to a dataset of sampled cells $\mathcal{S}$.
Common choices are the Bayesian Information Criterion (BIC) or the Akaike Information Criterion (AIC), which use $\kappa=\log(n_1)$ and $\kappa=2$, respectively. 
In particular, under random subsampling of empty cells of moderately large contingency tables, we consider the zero-deflated likelihood $\ell^{\text{zdpoi}}_{\mathcal{S}}$ in equation~\eqref{eq:sample-likelhood}, whereas for huge contingency tables we use the degenerate logistic likelihood $\ell_{\mathcal{S}}^{\text{dlogit}}$ in equation~\eqref{eq:dlogit_lik}.  Regardless of the number of empty cells sampled, the effective sample size remains $n_1=\sum_{\bx}n_{\bx}$, when conditioning on the total number of counts. The algorithm stops when no action results in an improvement of the information score.  
\textcolor{black}{As we consider the case of high-dimensionality, we induce a layer of stochasticity to this greedy stepwise procedure via hill-climbing: at any iteration, the algorithm performs either the action that improves the current score the most or the one that improves the score $N$ steps earlier.}

We evaluate this approach via a simulation study. In particular, we consider the same setting as before, with $p=13$ variables having $3$ levels each and a two-way interaction model. However, we now set a banded structure for the underlying conditional independence graph. As before, we draw all non-zero $\lambda$ effects  from a $\rm{Beta}$ distribution and set the intercept so as to have a sparse contingency table. In particular, the selected $\lambda$ parameters  lead on average to only $0.25\%$ of non-empty cells of the full contingency table. 
\begin{figure}[t]
    \centering
    \begin{minipage}[b]{0.45\linewidth}
        \centering        \includegraphics[width=\linewidth]{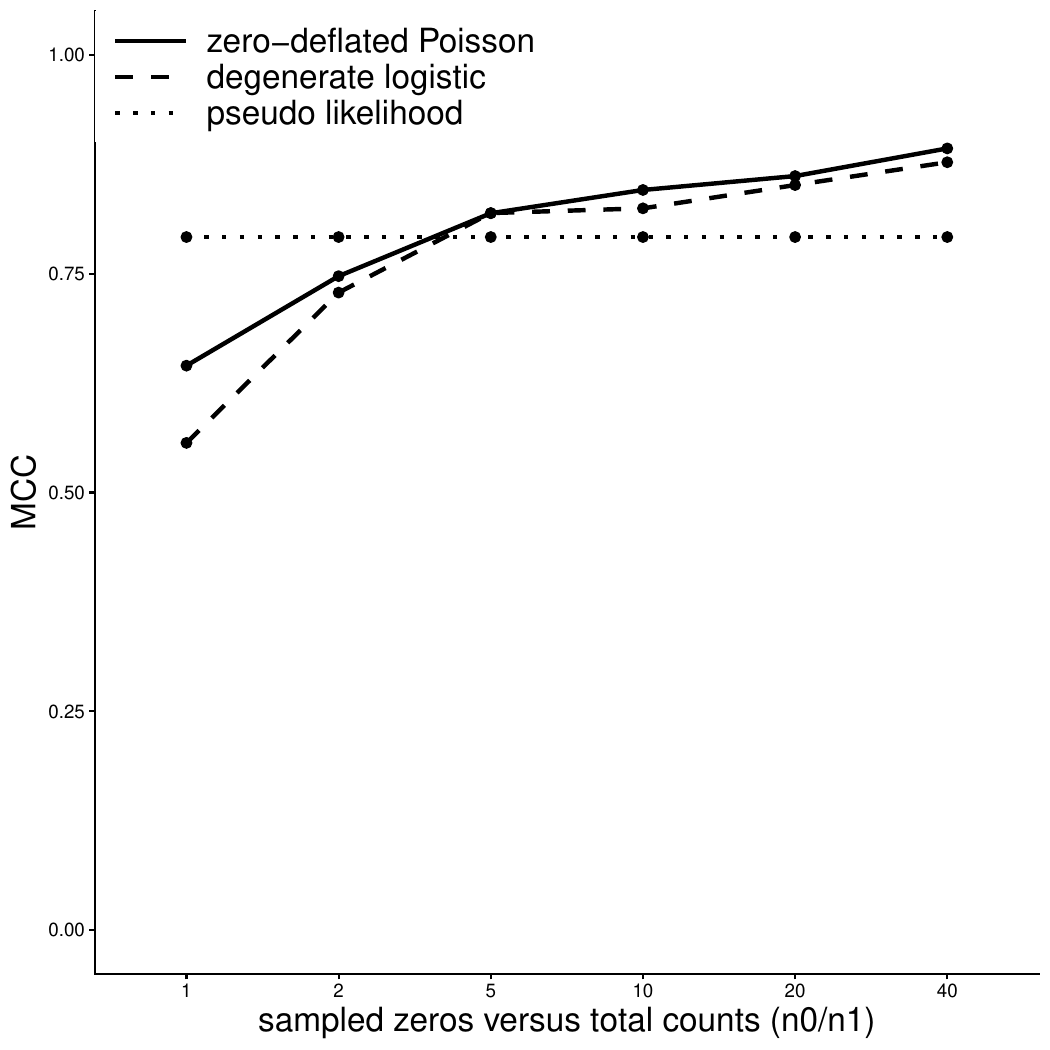}\\
(a)    
\end{minipage}
    \hfill
    \begin{minipage}[b]{0.45\linewidth}
        \centering
        \includegraphics[width=\linewidth]{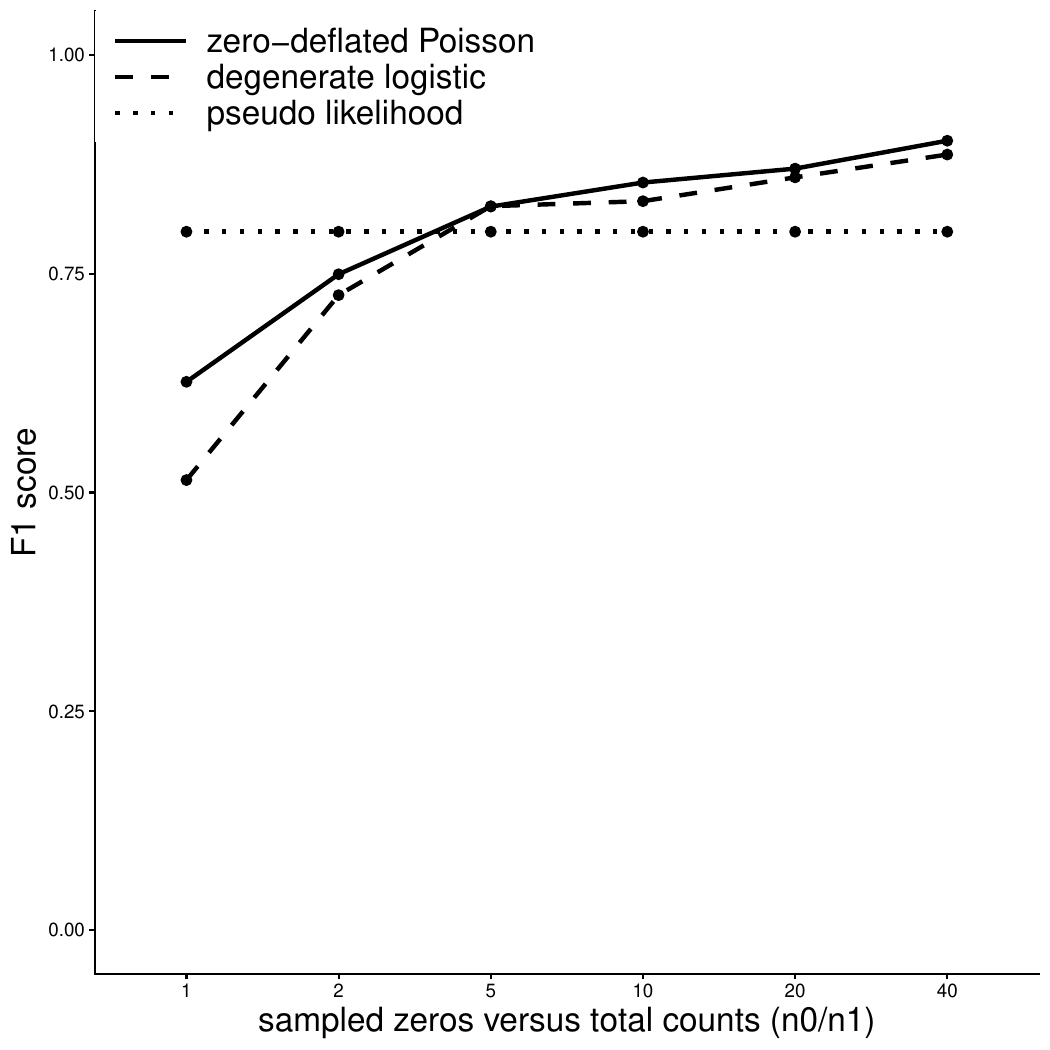}\\
(b)
    \end{minipage}  \hfill
    \caption{Simulation with $p=13$ categorical variables with 3 levels each   from a two-way model with a sparse banded structure and varying percentages of sampled zeros. Model selection via  a BIC stepwise procedure based either on the zero-deflated Poisson or the degenerate logistic likelihood is evaluated  according to (a) Matthews Correlation Coefficient and (b) $F_1$-score of the optimal model, and compared with the pseudo likelihood approach of \cite{dobra18}. For each setting, the  measures are averaged across $10$ replicates.
}
    \label{fig:modsel}
\end{figure}
For each scenario and each sampled dataset, we perform a graph-based stepwise procedure, where an edge is added or removed at each step based on the BIC of the resulting loglinear graphical model \textcolor{black}{and using $N=10$ steps for hill-climbing}. Estimation of $\boldsymbol{\lambda}$ and BIC calculation are performed both using the Poisson likelihood conditional on a random sample of zeros (section~\ref{sec:mediumlarge}) and  logistic regression from a nested case-control sampling scheme (section~\ref{sec:huge}). Figure~\ref{fig:modsel} evaluates the performance in terms of recovery of the true banded structure, that is detection of the non-zero $\lambda$ effects. Both the Matthews Correlation Coefficient (a) and the $F_1$ score (b) show a good recovery of the underlying graph already for a number of sampled zeros that are only an order of a magnitude higher than the total counts. 

\textcolor{black}{We finally include a comparison of the stepwise model selection procedure with the pseudo-likelihood approach of \cite{dobra18}, which explores the space of graphs by replacing the full multinomial likelihood with a pseudo-likelihood decomposed into the distribution of each node given its neighbours. As this approach does not model the loglinear effects, it can be followed by one of the methods described in this paper for parameter estimation. In particular, depending on the size of the contingency table, the IRWLS algorithm of section~\ref{sec:mediumlarge} or the degenerate logistic approach of section~\ref{sec:huge} can be used to fit the loglinear graphical model corresponding to the graph selected by a pseudo-likelihood approach. Since the implementation of the method in the R package \texttt{BDgraph} is only for multivariate binary data, we implement the approach for generic multivariate categorical data. Figure~\ref{fig:modsel} shows how the  approach has a superior performance in the case of only one or two controls per case, while looses accuracy as the percentage of sampled zeros increases, when both the zero-deflated Poisson and the logistic likelihood provide a better description of the true generative process. 
}

\section{Inferring cultural networks from survey data}\label{sec:realdata}

In social science, surveys are often collected to monitor the views of the population on important topics, with a number of international surveys that have been running on a regular basis for many years. Recent studies have used this rich source of data to describe the cultural values of a nation, to quantify the distances between national cultures or to detect possible trends and changes in views and attitudes over time, e.g., \cite{acemoglu24,bertrand23}. Among these quantitative studies, some  have emphasized the fact that  accounting for the dependence structure of the different cultural dimensions leads to a more comprehensive understanding of a national culture \citep{debenedictis23,vinciotti24}. In this section, we follow this line of research, and show how the computational methods presented in this paper allow us to perform loglinear modelling of survey data. 

Survey data present a typical example of a large contingency table. In these data, the multivariate categorical response of an individual to a survey  represents a count in the contingency table. Thus, a moderate number of questions, each with a moderate number of levels, are sufficient to generate extremely large contingency tables.
For the analysis, we consider in particular the data from the General Social Survey \citep{GSS2024}. Similar to the study of \citep{bertrand23}, we consider \textcolor{black}{$69$ questions} that assess the characteristics and views of the American population on the following seven broad topics:
\begin{enumerate}
\item \textit{Civil liberties}: allow atheists to teach (\texttt{colath}); allow communists to teach (\texttt{colcom}); allow racists to teach (\texttt{colrac}); allow atheists' books in library (\texttt{libath}); allow communists' books in library (\texttt{libcom}); allow racists' books in library (\texttt{librac}); allow atheists to speak (\texttt{spkath}); allow communists to speak (\texttt{spkcom}); allow racists to speak (\texttt{spkrac}).
\item \textit{Confidence}: confidence in military (\texttt{conarmy}); confidence in business (\texttt{conbus}); confidence in organized religion (\texttt{conclerg}); confidence in education (\texttt{coneduc}); confidence in executive branch (\texttt{confed}); confidence in financial institutions (\texttt{confinan}); confidence
in US Supreme Court (\texttt{conjudge}); confidence in organized labor (\texttt{conlabor}); confidence in congress (\texttt{conlegis}); confidence in medicine (\texttt{conmedic}); confidence in the press (\texttt{conpress}); confidence in scientific community (\texttt{consci}); confidence in TV (\texttt{contv}).
\item \textit{Government spending}: foreign aid (\texttt{nataid}); military \& defense (\texttt{natarms}); solving problems of large cities (\texttt{natcity}); halting crime rate (\texttt{natcrime}); dealing with drug addiction (\texttt{natdrug}); education (\texttt{nateduc}); environment (\texttt{natenvir}); welfare (\texttt{natfare}); health care (\texttt{natheal}); space exploration programs (\texttt{natspac}); income tax too high/adequate/too low (\texttt{tax}).
\item \textit{Law enforcement and gun control}: courts dealing with criminals (\texttt{courts}); should marijuana be legal (\texttt{grass}); approve of police striking citizens if: citizen said vulgar things (\texttt{polabuse}); citizen attempted to escape custody (\texttt{polescap}); citizen questioned as murder suspect (\texttt{polmurdr}); ever approve of police striking citizen (\texttt{polhitok}); favour/oppose death penalty for murder (\texttt{cappun}); favour/oppose gun permits (\texttt{gunlaw}); have gun at home (\texttt{owngun}).
\item \textit{Life, life outlook, and trust}: should aged live with their children (\texttt{aged}); afraid to walk at night in neighborhood (\texttt{fear}); opinion of how people get ahead (\texttt{getahead}); general happiness (\texttt{happy}); condition of health (\texttt{health}); people helpful
or looking out for themselves (\texttt{helpful}); any opposite race in neighborhood (\texttt{raclive}); if rich, continue or stop working (\texttt{richwork}); job satisfaction (\texttt{satjob}); can people be trusted (\texttt{trust}).
\item \textit{Marriage, sex, and abortion}: approve of legal abortion if: strong chance of serious defect (\texttt{abdefect}); woman's health seriously endangered (\texttt{abhlth}); married-wants no more children (\texttt{abnomore}); low income-cannot afford more children (\texttt{abpoor}); pregnant as result of rape (\texttt{abrape}); not married (\texttt{absingle}); opinion about homosexual sex relations (\texttt{homosex}); sexual relations before marriage (\texttt{premarsx}); extramarital sex (\texttt{xmarsex}); divorce laws (\texttt{divlaw}); porn laws (\texttt{pornlaw}); seen X-rated movie in the last year (\texttt{xmovie}).
\item \textit{Politics and religion}: political party affiliation (\texttt{partyid}, with levels recoded  to: democratic, independent, republican);  how often attend religious services (\texttt{attend}, with levels recoded to: none, rarely, often, very often); religion denomination (\texttt{relig}, with levels recoded to: protestant, catholic, other, none);  belief in life after death (\texttt{postlife}).
\end{enumerate}

With the exception of 2021, when some questions were not included, we use data from eight survey waves conducted between 2008 and 2024, focusing on respondents aged 20 to 64. The survey is administered in three separate ballots, so each individual question is answered by roughly two-thirds of participants. As a result, a substantial share of the missing data is structural. We address this by applying a multivariate imputation approach based on classification trees, implemented in the R package \texttt{mice} \citep{mice}. The remaining missing observations --  approximately 1.8\% of the data -- correspond to respondents who did not know or declined to answer. These are treated as an additional category for each question. Consequently, all variables have more than two levels, and these levels are unordered, even for variables whose observed categories have a natural ordering.  After recoding three variables with many categories (party affiliation, religion, and church attendance), the maximum number of levels per variable is five. The resulting \textcolor{black}{69-dimensional contingency table contains approximately $6.6 \times 10^{38}$ cells}, of which only $15,014$ have counts in them, implying that nearly all cells are empty.

 \begin{figure}[h!]
\centering
\begin{minipage}[b]{\linewidth}
\centering
\includegraphics[width=\textwidth]{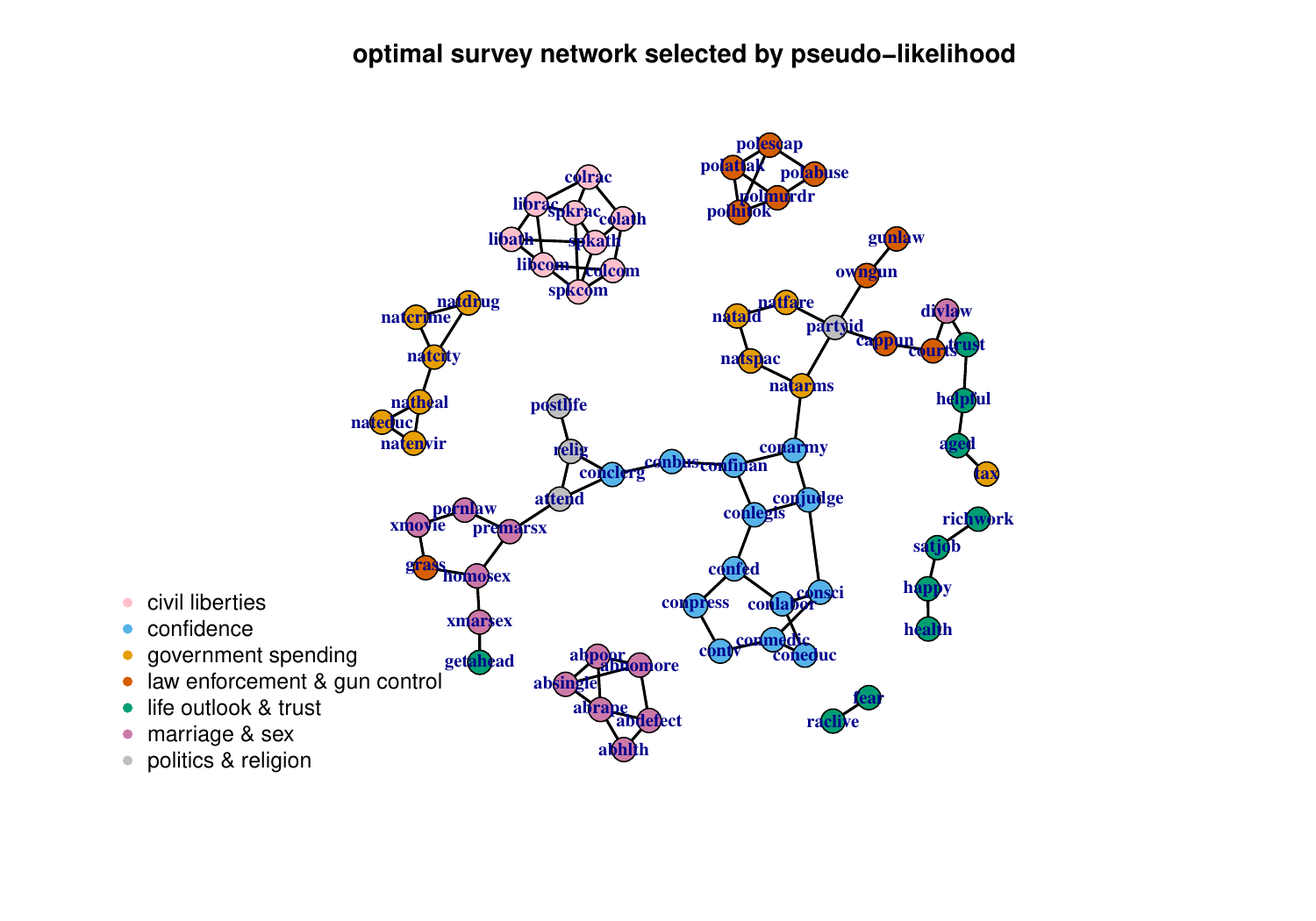} \\
(a)
\end{minipage}
\begin{minipage}[b]{0.49\linewidth}
\centering
\includegraphics[width=\linewidth]{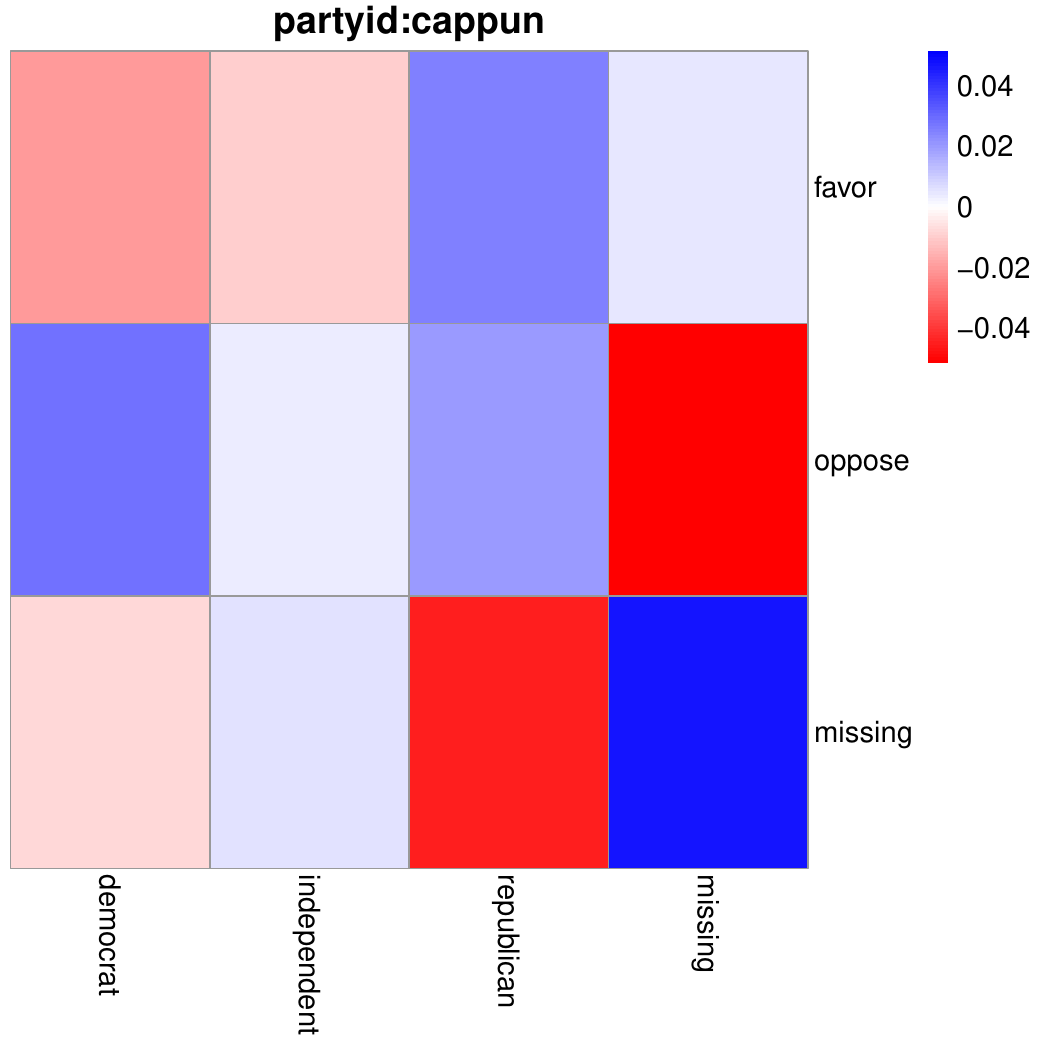}\\
(b)
\end{minipage}
\begin{minipage}[b]{0.49\linewidth}
\centering
\includegraphics[width=\linewidth]{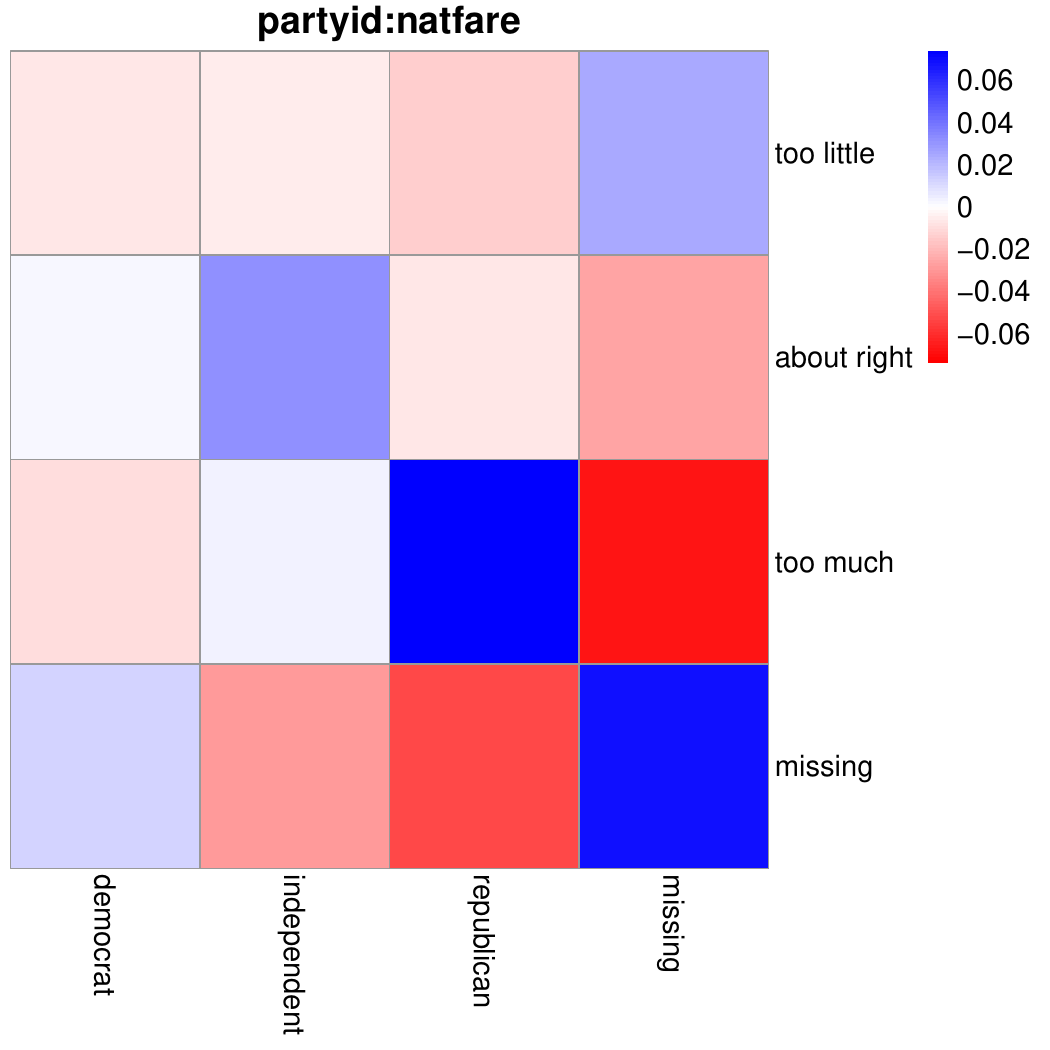}\\
(c)
\end{minipage}
\caption{(a) Optimal network on survey data, found by a stochastic search based on the multinomial pseudo-likelihood. (b)-(c) Estimated two-way effects for two of the interactions in the selected loglinear graphical model: republicans tend to be more in favour of death penalty and to think that government spending on welfare is too high compared to democrats and independents.}
\label{fig:survey}
\end{figure}
\textcolor{black}{We randomly generate $5$ controls for each case, so $75070$ empty cells, and use the degenerate logistic approach to fit the loglinear graphical model selected by a pseudo-likelihood approach. For parameter estimation, we use a ridge regularization with a tuning parameter selected by generalized cross-validation.  Figure~\ref{fig:survey}a shows the conditional independence graph of the selected model, which has $52$ two-way interactions and $12$ three-way interactions, besides the main effects.} 
 The colouring of the nodes corresponds to the seven broad topics described above. As expected, there are some evident dependences among answers to questions related to the same topic. However, the graph shows also dependences among different dimensions of culture, consolidating the view that a national culture consists of a number of inter-connected dimensions  \citep{debenedictis23}. 
\textcolor{black}{ Figures~\ref{fig:survey}(b) and (c) visualize two of the dependences found with the node \texttt{partyid}. This node reports the political affiliation of survey respondents. The analysis shows how this is associated with the views of people about the death penalty (\texttt{cappun}) and the government spending on improving the nation's welfare (\texttt{natfare}). Heatmaps of the two-way effects, estimated with a sum-to-zero constraint, show how, compared to democrats and independents,  republicans tend to be more in favour of the death
penalty and to think that government spending on welfare is too high.}

\section{Conclusion}\label{sec:conclusion}
Extremely large contingency tables are collected in many application areas. While efficient approaches exist for learning structural dependences from these data \citep{dobra18},   parameter estimation of loglinear graphical models from huge contingency tables remains computationally prohibitive.

\textcolor{black}{In this paper, we have  shown how statistical inference of loglinear models from sparse and high-dimensional contingency tables can be conducted via subsampling on the set of the empty cells. In particular, we propose a zero-deflated Poisson approach  for moderately size contingency tables and nested case-control multinomial sampling for huge contingency tables. Both approaches can be easily augmented with a ridge regularization for ill-conditioned problems, where the maximum likelihood estimators do not exist. We discuss properties of the estimators  of the proposed methods, together with parameter estimation, model selection and computational complexity, and evaluate these via a simulation study. }
An illustration is presented on survey data from the General Social Survey, where the nested case-control sampling approach is used to recover and quantify the dependence structure among different dimensions of culture from a huge contingency table. 

\section*{Code availability}
\textcolor{black}{The \texttt{R} script to replicate the simulation study and the survey data analysis is available from
\url{https://github.com/veronicavinciotti/loglin}.}

\section*{Acknowledgement}
Veronica Vinciotti acknowledges funding from the the European Union
- Next Generation EU, Mission 4 Component 2 - CUP C53D23002580006 (MUR-PRIN grant
2022SMNNKY). We thank the anonymous reviewers for their constructive feedback and insightful suggestions, which greatly improved the
quality of the manuscript.
\bibliography{biblio}

@article{birch1963maximum,
  author  = {Birch, M. W.},
  title   = {Maximum Likelihood Estimation of $\mathcal{L}$-Relation Table Probabilities},
  journal = {Journal of the Royal Statistical Society: Series B (Methodological)},
  year    = {1963},
  volume  = {25},
  number  = {1},
  pages   = {220--233},
  doi     = {10.1111/j.2517-6161.1963.tb00502.x}
}

@book{bishop1975discrete,
  author    = {Bishop, Yvonne M. M. and Fienberg, Stephen E. and Holland, Paul W.},
  title     = {Discrete Multivariate Analysis: Theory and Practice},
  publisher = {MIT Press},
  address   = {Cambridge, MA},
  year      = {1975},
  isbn      = {978-0262021135}
}

@article{varin2011overview,
  title={An overview of composite likelihood methods},
  author={Varin, Cristiano and Reid, Nancy and Firth, David},
  journal={Statistica Sinica},
  pages={5--42},
  year={2011},
  publisher={JSTOR}
}

@article{behrouzi2019detecting,
  title={Detecting epistatic selection with partially observed genotype data by using copula graphical models},
  author={Behrouzi, Pariya and Wit, Ernst C},
  journal={Journal of the Royal Statistical Society Series C: Applied Statistics},
  volume={68},
  number={1},
  pages={141--160},
  year={2019},
  publisher={Oxford University Press}
}

@article{mohammadi17,
author = {Mohammadi, R. and Abegaz, F. and van den Heuvel, E. and Wit, E. C.},
title = {Bayesian modelling of {Dupuytren} disease by using {Gaussian} copula graphical models},
journal = {Journal of the Royal Statistical Society: Series C (Applied Statistics)},
volume = {66},
number = {3},
pages = {629-645},
year = {2017}
}

@article{debenedictis23,
    author = {De Benedictis, Luca and Rondinelli, Roberto and Vinciotti, Veronica},
    title = {Cultures as networks of cultural traits: a unifying framework for measuring culture and cultural distances},
    journal = {Journal of the Royal Statistical Society Series A: Statistics in Society},
    volume = {186},
    number = {3},
    pages = {264-293},
    year = {2023}
}

@article{vinciotti24,
    author = {Vinciotti, Veronica and De Benedictis, Luca and Wit, Ernst C},
    title = {Joint modelling of national cultures accounting for within and between-country heterogeneity},
    journal = {Journal of the Royal Statistical Society Series A: Statistics in Society},
    volume = {188},
    number = {4},
    pages = {1231-1245},
    year = {2024}
}

@article{acemoglu24,
    author = {Acemoglu, Daron and Ajzenman, Nicolás and Aksoy, Cevat Giray and Fiszbein, Martin and Molina, Carlos},
    title = {(Successful) Democracies Breed Their Own Support},
    journal = {The Review of Economic Studies},
    volume = {92},
    number = {2},
    pages = {621-655},
    year = {2024}
}

@Article{mice,
    title = {{mice: Multivariate Imputation by Chained Equations in R}},
    author = {Stef {van Buuren} and Karin Groothuis-Oudshoorn},
    journal = {Journal of Statistical Software},
    year = {2011},
    volume = {45},
    number = {3},
    pages = {1-67}
  }

@misc{GSS2024,
title = {{General Social Survey 1972-2024. Data accessed from the GSS Data Explorer} website at gssdataexplorer.norc.org.
},
year = {2025},
author = {Davern, Michael and Bautista, Rene and Freese, Jeremy and Herd, Pamela and Morgan, Stephen L.}  }

@article{fienberg12,
  author  = {Fienberg, Stephen E. and Rinaldo, Alessandro},
  title   = {Maximum likelihood estimation in log-linear models},
  journal = {The Annals of Statistics},
  year    = {2012},
  volume  = {40},
  doi     = {10.1214/12-aos986}
}

@article{nardi12,
title = {The log-linear group-lasso estimator and its asymptotic properties},
journal = {Bernoulli},
volume = {18},
number = {3},
pages = {945 -- 974},
year = {2012},
author = {Yuval Nardi and Alessandro Rinaldo}
}

@article{aliverti22,
author = {Emanuele Aliverti and David B. Dunson},
title = {{Composite mixture of log-linear models with application to psychiatric studies}},
volume = {16},
journal = {The Annals of Applied Statistics},
number = {2},
publisher = {Institute of Mathematical Statistics},
pages = {765 -- 790},
year = {2022}
}

@book{agresti12,
  address = {Hoboken, NJ, USA},
  author = {Agresti, Alan},
  publisher = {Wiley},
  title = {Categorical Data Analysis, 3rd Edition},
  year = 2012
}

@book{roverato17, 
address={Cambridge}, 
series={SemStat Elements}, 
title={Graphical Models for Categorical Data}, 
publisher={Cambridge University Press}, author={Roverato, Alberto}, 
year={2017}, collection={SemStat Elements}
}

@book{lauritzen96,
	Address = {Oxford},
	Author = {S. L. Lauritzen},
	Publisher = {Oxford University Press},
	Title = {Graphical Models},
	Year = {1996}}

@Article{glmnet,
    title = {Regularization Paths for Generalized Linear Models via Coordinate Descent},
    author = {Jerome Friedman and Trevor Hastie and Robert Tibshirani},
    journal = {Journal of Statistical Software},
    year = {2010},
    volume = {33},
    number = {1},
    pages = {1--22}
  }

@article{wang22,
  title={Maximum sampled conditional likelihood for informative subsampling},
  author={Haiying Wang and Jae Kwang Kim},
  journal={Journal of Machine Learning Research},
  year={2022},
  volume={23},
  issue={1}
}

@book{hoisgaard12,
    title = {Graphical Models with {R}},
    author = {S{\o}ren H{\o}jsgaard and David Edwards and Steffen Lauritzen},
    publisher = {Springer},
    address = {New York},
    year = {2012}}

@book{McCullagh89,
  address = {London},
  author = {McCullagh, P. and Nelder, J. A.},
  location = {London, UK},
  publisher = {Chapman \& Hall / CRC},
  title = {Generalized Linear Models},
  year = {1989}
}

@Article{bdgraph,
    author = {Reza Mohammadi and Ernst C Wit},
    title = {{BDgraph}: An {R} Package for {B}ayesian Structure Learning in Graphical Models},
    journal = {Journal of Statistical Software},
    year = {2019},
    volume = {89},
    number = {3},
    pages = {1--30}
  }

@article{dobra18,
author = {Adrian Dobra and Reza Mohammadi},
title = {{Loglinear model selection and human mobility}},
volume = {12},
journal = {The Annals of Applied Statistics},
number = {2},
pages = {815 -- 845},
year = {2018}
}

@article{dahinden10,
author = {Dahinden, Corinne and Kalisch, Markus and B\"uhlmann, Peter},
title = {Decomposition and Model Selection for Large Contingency Tables},
journal = {Biometrical Journal},
volume = {52},
number = {2},
pages = {233 -- 252},
year = {2010}
}

@article{bertrand23,
Author = {Bertrand, Marianne and Kamenica, Emir},
Title = {Coming Apart? {Cultural} Distances in the {United States} over Time},
Journal = {American Economic Journal: Applied Economics},
Volume = {15},
Number = {4},
Year = {2023},
Pages = {100--141},
}

@article{Breslow1974,
	author  = {Breslow, Norman E.},
	title   = {Covariance analysis of censored survival data},
	journal = {Biometrics},
	volume  = {30},
	number  = {1},
	pages   = {89--99},
	year    = {1974}
}

@article{Prentice1978,
	author  = {Prentice, Ross L. and Breslow, Norman E.},
	title   = {Retrospective studies and failure time models},
	journal = {Biometrika},
	volume  = {65},
	number  = {1},
	pages   = {153--158},
	year    = {1978}
}

@article{Thomas1977,
	author  = {Thomas, Duncan C.},
	title   = {Addendum to ``{M}ethods of cohort analysis: Appraisal by application to
		asbestos mining'' by {L}iddell, {M}c{D}onald and {T}homas},
	journal = {Journal of the Royal Statistical Society: Series A (General)},
	volume  = {140},
	number  = {4},
	pages   = {483--485},
	year    = {1977}
}

@article{Oakes1981,
	author  = {Oakes, David},
	title   = {Survival times: aspects of partial likelihood (with discussion)},
	journal = {International Statistical Review},
	volume  = {49},
	number  = {3},
	pages   = {235--264},
	year    = {1981}
}

@article{Goldstein1992,
	author  = {Goldstein, Larry and Langholz, Bryan},
	title   = {Asymptotic theory for nested case-control sampling in the {C}ox regression model},
	journal = {The Annals of Statistics},
	volume  = {20},
	number  = {4},
	pages   = {1903--1928},
	year    = {1992},
	doi     = {10.1214/aos/1176348895}
}

@article{Langholz1996,
	author  = {Langholz, Bryan and Goldstein, Larry},
	title   = {Risk set sampling in epidemiologic cohort studies},
	journal = {Statistical Science},
	volume  = {11},
	number  = {1},
	pages   = {35--53},
	year    = {1996}
}

@book{Therneau2000,
	author    = {Therneau, Terry M. and Grambsch, Patricia M.},
	title     = {Modeling Survival Data: {E}xtending the {C}ox Model},
	publisher = {Springer},
	address   = {New York},
	year      = {2000},
	isbn      = {0-387-98784-3}
}
\end{document}